\documentclass[reprint,amsmath,amssymb,aps,floatfix,superscriptaddress,prx]{revtex4-2}

\usepackage{graphicx}
\usepackage{float}
\usepackage[colorlinks=true,linkcolor=blue,citecolor=blue,urlcolor=blue]{hyperref}
\usepackage{listings}
\usepackage[utf8]{inputenc}
\usepackage[english]{babel}

\graphicspath{{images/}} 

\begin{document}

\title{Deep learning estimation of complex reverberant wave fields with a programmable metasurface}

\author{Benjamin W. Frazier}
\email{frazibw1@umd.edu}
\affiliation{Institute for Research in Electronics and Applied Physics, University of Maryland, College Park, MD 20742, USA}
\affiliation{Department of Electrical and Computer Engineering, University of Maryland, College Park, MD 20742, USA}
\affiliation{Johns Hopkins University Applied Physics Laboratory, Laurel, MD 20723, USA}

\author{Thomas M. Antonsen, Jr.}
\affiliation{Institute for Research in Electronics and Applied Physics, University of Maryland, College Park, MD 20742, USA}
\affiliation{Department of Electrical and Computer Engineering, University of Maryland, College Park, MD 20742, USA}
\affiliation{Department of Physics, University of Maryland, College Park, MD 20742, USA}

\author{Steven M. Anlage}
\affiliation{Department of Electrical and Computer Engineering, University of Maryland, College Park, MD 20742, USA}
\affiliation{Department of Physics, University of Maryland, College Park, MD 20742, USA}
\affiliation{Quantum Materials Center, University of Maryland, College Park, MD 20742, USA}

\author{Edward Ott}
\affiliation{Institute for Research in Electronics and Applied Physics, University of Maryland, College Park, MD 20742, USA}
\affiliation{Department of Electrical and Computer Engineering, University of Maryland, College Park, MD 20742, USA}
\affiliation{Department of Physics, University of Maryland, College Park, MD 20742, USA}
\date{\today}

\begin{abstract}
Electromagnetic environments are becoming increasingly complex and congested, creating a growing challenge for systems that rely on electromagnetic waves for communication, sensing, or imaging, particularly in reverberating environments. The use of programmable metasurfaces provides a potential means of directing waves to optimize wireless channels on-demand, ensuring reliable operation and protecting sensitive electronic components. Here we introduce a technique that combines a deep learning network with a binary programmable metasurface to shape waves in complex reverberant electromagnetic environments, in particular ones where there is no direct line of sight. We applied this technique for wavefront reconstruction and control, and accurately determined metasurface configurations based on measured system scattering responses in a chaotic microwave cavity. The state of the metasurface that realizes desired electromagnetic wave field distribution properties was successfully determined even in cases previously unseen by the deep learning algorithm. Our technique is enabled by the reverberant nature of the cavity, and is effective with a metasurface that covers only $\sim$1.5\% of the total cavity surface area.
\end{abstract}

\maketitle

\section{\label{sec:intro}Introduction}
Highly scattering environments scramble electromagnetic waves, producing interference among the multiple paths between source and receiver. The resulting spatio-temporal fluctuations can seriously degrade imaging, sensing, and communication systems at microwave and optical wavelengths, disrupting operation or even damaging sensitive components. Additional emissions in these environments, whether from unintentional coupling between components or from an intentional electromagnetic attack, can have serious consequences. Future smart radio environments are envisioned to handle such dynamic conditions, adapting on-the-fly to optimize a given wireless channel through a spatial light modulator (SLM) \cite{subrtSmartRadio2012,direnzoSmartRadioAI,direnzoSmartRadio}. Intelligently controlling wave fields in the presence of multi-path reflections is therefore a critical factor for enabling smart radio environments. In addition, an intelligent and self-adaptive approach will benefit applications such as micromanipulation of objects in complex scattering environments \cite{horodynskiOptimal2020}, and time reversal mirrors that can selectively focus a wavefront or enhance communication system performance \cite{frazierNonlinearTimeReversal2013, xiaoFocusingWavesArbitrary2016}. A necessary step along this path is to identify approaches for wavefront reconstruction, i.e., determining the configuration of the SLM that accurately produces a given scattering response, that work in complicated scattering environments. 

In optics, SLMs have been used to control waves under strong scattering conditions for some time. Applications range from focusing through general disordered media \cite{vellekoopFocusingCoherentLight2007, moskControl2012} to sophisticated biomedical imaging instruments that fall under the umbrella of adaptive optics \cite{boothAberrationsAdaptiveOptics2015,colliniAdaptiveOpticsMicrospectrometer2019}. In the last several years, spatial microwave modulators in the form of programmable metasurfaces have also become widely available. Programmable electromagnetic metasurfaces are metamaterial sheets that can modify their local surface impedance over unit cells (meta-atoms) that have a sub-wavelength characteristic size. They have emerged as powerful tools for shaping waves inside complex microwave cavities \cite{dupreWaveFieldShapingCavities2015,kainaShapingComplexMicrowave2015,delhougneSpatiotemporalWaveFront2016,delhougneOptimalMultiplexingSpatially2020,grosTuningRegularCavity2020, imaniPerfectAbsorptionDisordered2020,delhougneImplementingNonUniversalFeaturesPhysRevE2020,frazierWavefrontShaping2020,delHougneOnDemand2020}. 

Metasurfaces are not limited to shaping only electromagnetic waves. In seismology, control over surface acoustic waves has been demonstrated using metasurfaces made of elastic metamaterials for Love waves (horizontally polarized) \cite{palermoControl2018} and Rayleigh waves (containing both longitudinal and transverse motion) \cite{heSeismic2020}. In the case of quantum waves, a metasurface created from an array of trapped neutral atoms was used to manipulate light at the quantum level \cite{bekensteinQuantum2020}. While the underlying physics of these metasurfaces is vastly different, the overall operation and process of wave interaction is essentially the same, implying that strategies for wavefront shaping in one domain can be readily adapted to another.

Wavefront shaping techniques with metasurfaces have been well studied; however, control in complex reverberating environments still relies on simple, online brute force optimization methods. While these approaches work, they require a large number of iterations to reach convergence, are rarely guaranteed to achieve a global minimum, and can produce undesirable scattering configurations through the intermediate steps of the optimization process. Wavefront reconstruction, or estimating the wavefront in the basis of metasurface commands, is therefore a critical capability for enabling practical wavefront control applications.

\begin{figure*}
\includegraphics[width = 17.2cm]{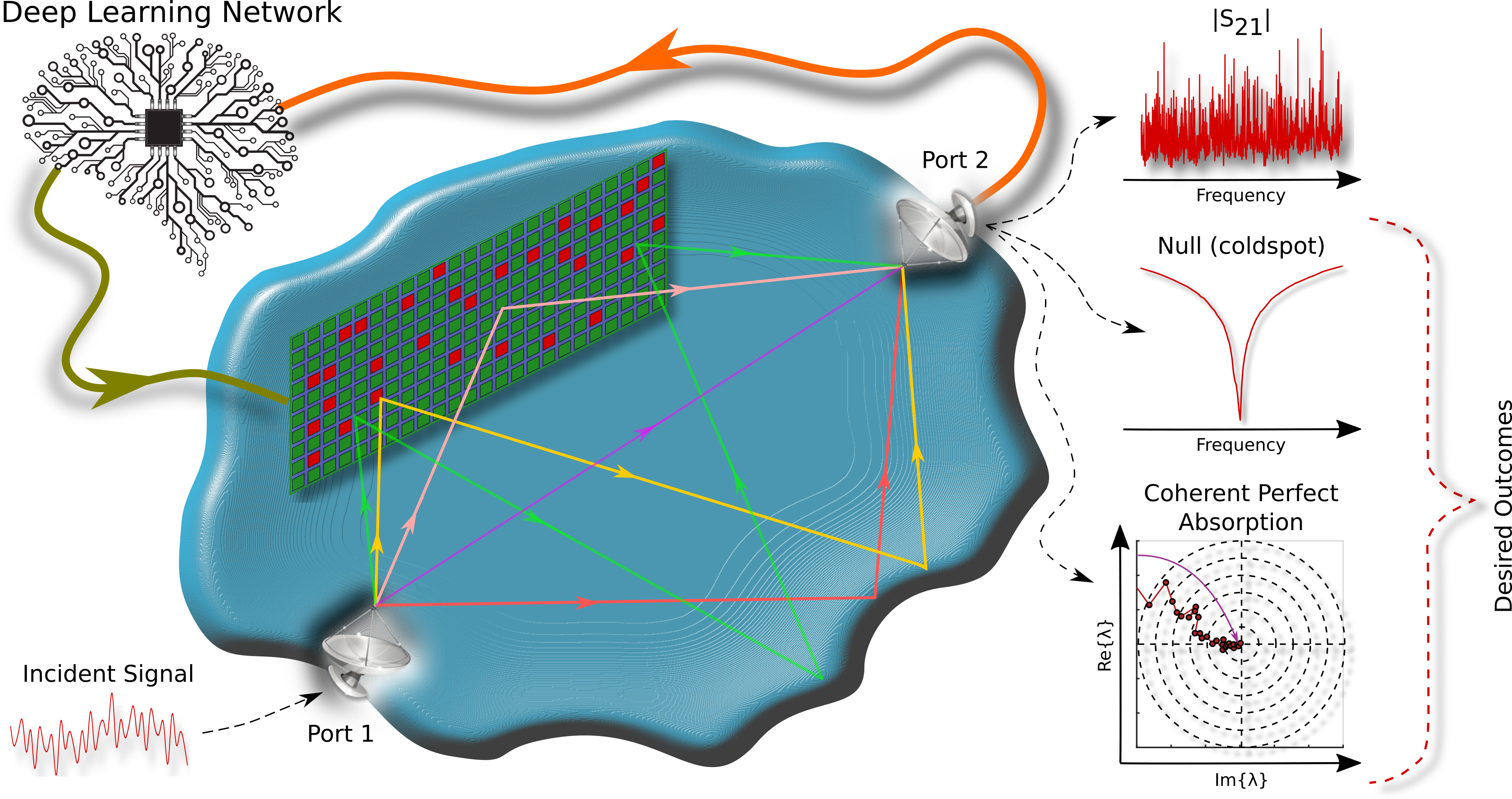}
\caption{\label{fig:overview} \bf Conceptual view of a deep learning enabled programmable metasurface in a complex electromagnetic environment. \normalfont Constructive and destructive interference between multiple propagation paths in a reverberating environment induces randomness in the scattering parameters and scrambles electromagnetic waves that are injected at Port 1. A reconfigurable metasurface is used to tune the interference to create cold spots for protection of sensitive electronic components, realize coherent perfect absorption states for long range wireless power transfer, or unscramble the output fields to enable smart radio environments. The metasurface, along with a sensing antenna at Port 2, is coupled with a deep learning network that provides control. Measurements are used as training data, enabling the network to determine the control settings of the metasurface, and allowing the system to adapt to changing environmental conditions on-the-fly. The metasurface is shown here as large relative to the cavity. In our configuration however, the metasurface is much smaller, covering only $\sim$1.5\% of the total surface area of the cavity.}
\end{figure*}

In this Article, we use a binary programmable metasurface to shape radio frequency electromagnetic waves inside a chaotic microwave cavity, and present a deep learning network that solves the wavefront reconstruction problem, enabling real-time operation once trained. We show that the deep learning network achieves an accuracy exceeding 99\%. This high success rate is achieved with a limited amount of training data, requiring the collection of far fewer sets than the number of possible combinations of metasurface commands. A conceptual view of our technique is given in Fig. \ref{fig:overview}. The metasurface is placed in a reverberant scattering environment, with a signal injected at Port 1 and the resulting field measured at a specific point of interest (Port 2). The environment is defined by irregular walls and inclusions and is probed by waves with wavelengths much smaller than the characteristic dimension of the enclosure.

We emphasize that our method is enabled by the use of a reverberant environment, which allows the metasurface to interact with multiple ray trajectories, often more than once. A reverberant environment provides two major capabilities that are not present in non-reverberant environments: 1) the ability to control the distribution of wave fields at arbitrarily chosen locations inside the cavity is enhanced. This allows the use of relatively small metasurfaces, e.g., in our configuration, the metasurface covers only $\sim$1.5\% of the total surface area of the cavity; and 2) the requirement on establishing a line-of-sight path between the metasurface and the ports is removed, which allows the location of the metasurface to be arbitrarily chosen, increasing the flexibility and versatility of the approach. We anticipate that realization of this concept will help usher in the new era of smart radio environments, as well as allow on-demand creation of microwave cold spots to protect sensitive electronic components and coherent perfect absorption states for wireless power transfer.

\section{\label{sec:deepwavefront}Wavefront Control in Reverberant Environments}
Microwave experiments have shown that programmable metasurfaces can provide fine control over the scattering parameters of a cavity, with the most recent work demonstrating perfect absorption \cite{imaniPerfectAbsorptionDisordered2020} and coherent perfect absorption \cite{frazierWavefrontShaping2020,delHougneOnDemand2020} states inside the cavity. The relationship between metasurface commands and cavity scattering parameter responses is extremely complicated (there are $10^{18}$ possible configurations of the metasurface in our case). Therefore, optimization of the metasurface is typically handled through brute force trial and error or stochastic search algorithms \cite{vellekoopPhase2008,dorrerDirectBinarySearch2018,frazierWavefrontShaping2020}. As discussed in Appendix \ref{app:wavefront_reconstruction_control}, rapid and accurate wavefront reconstruction techniques that solve the inverse problem between measurements and metasurface commands are necessary to realize practical intelligent wavefront shaping systems. Conventional methods fall apart in complex scattering environments with binary metasurfaces; however, the inherent complexity makes it an ideal place to utilize deep learning. Ma et al. explored the use of deep learning networks with wave chaotic systems, demonstrating the ability to successfully distinguish between different types of wave chaotic cavities through the measured $S$-parameters \cite{maClassificationPredictionWave2019}. We now tackle a more difficult problem, identifying a set of metasurface commands required to achieve a specific wave scattering condition, even for cases where that set of commands has not been previously encountered.

Deep learning has been successfully used to design metasurfaces for wavefront shaping applications in both the photonic and microwave domains \cite{peurifoyANN2018,liuTrainingDeepNeural2018,nadellDeepLearningAccelerated2019,anDeepLearningApproach2019,qiuDeepLearningRapid2019,sajedianDoubledeepQlearningIncrease2019,yeungElucidatingBehaviorNanophotonic2020,wiechaDeepLearningMeets2020, jiangDeepNeuralNetworks2020,unniDeepConvolutionalMixture2020, mallCyclicalDeepLearning2020, anDeepLearningModeling2020}. However, most of the publications so far have focused on designing and arranging the individual unit cells of the metasurface for static use cases. Active deep learning approaches with programmable metasurfaces have been demonstrated for microwave imaging applications \cite{liIntelligentMetasurfaceImager2019,liMachinelearningReprogrammableMetasurface2019,delhougneLearnedIntegratedSensing2020, liLearnedSensing2020,delHougneRobustPosition2020,delHougneDeeplySubWavelength2021}. Li et al. used a two-bit coding metasurface to generate radiation patterns for a machine learning algorithm that detects and classifies human movement \cite{liIntelligentMetasurfaceImager2019, liMachinelearningReprogrammableMetasurface2019}. del Hougne et al. started with a pair of metasurfaces as a transmitter and receiver to feed a dense neural network that detects and classifies objects in a learned integrated sensing paradigm \cite{delhougneLearnedIntegratedSensing2020,liLearnedSensing2020}. Further research by this group used a dense neural network to classify the position of a scattering object inside a complex cavity with a metasurface acting as a coded aperture \cite{delHougneRobustPosition2020}; this work was recently extended to predict a continuous position with sub-wavelength precision \cite{delHougneDeeplySubWavelength2021}. 

These examples demonstrate how a programmable metasurface can enhance the processing power of a deep learning network for microwave imaging, but they do not leverage the deep learning network for wavefront reconstruction. This is a key component of intelligent wavefront shaping, which has so far been an underexplored area of research. Qian et al. used a simple dense network to enable cloaking of an object \cite{qianDeeplearningenabledSelfadaptiveMicrowave2020}, while Shan et al. used a 2D convolutional network to optimize the steering of multiple beams \cite{shanCodingProgrammableMetasurfaces2020}. Both cases utilize an idealized testing environment inside an anechoic chamber, where multi-path reflections from the environment are intentionally excluded. In addition, both cases are built around a propagation path with a direct reflection off the metasurface, which means that the metasurface interacts with virtually all ray trajectories from the source to the receiver.

As discussed in Appendix \ref{app:complexScattering}, a single propagation path eliminates redundancies from persistent short orbits \cite{hartEffectShortRay2009, yehExperimentalExaminationEffect2010}, reducing the measured correlation between metasurface configurations. These cases can be treated with more traditional system identification techniques or simple neural network models. When the metasurface is placed inside a complex reverberant scattering volume \cite{delHougneRobustPosition2020, delHougneDeeplySubWavelength2021}, determining the relationship between metasurface commands and scattering responses becomes substantially more difficult due to the presence of multiple scattering paths. A reverberant scattering system is qualitatively different from an open system, and is characterized by extreme sensitivity to initial conditions \cite{ottChaosDynamicalSystems2002,haakeQuantumSignaturesChaos2010}. This means accurate wavefront reconstruction must account for chaotic behavior and be sensitive to small environmental changes, as well as handle non-negligible large amplitude signal spikes that include phenomena such as rogue waves \cite{hohmannFreakWaves2010}. This difficulty is further compounded as we wish to optimize the metasurface response over a wide bandwidth or even over multiple separated bandwidths simultaneously.

The reverberating nature of the cavity enables operation with a smaller metasurface than would be possible in a non-reverberating environment. Longer reverberation times (lower cavity losses) mean that the rays will survive longer in the cavity, resulting in more reflections from scattering objects and more ray trajectories that interact with the metasurface, often multiple times. Longer reverberation times then provide the metasurface more flexibility in controlling constructive and destructive interference at the ports, allowing for larger relative changes when toggling metasurface states. We demonstrate this to be the case and show that the performance of the deep learning network degrades as the losses in the cavity increase because the metasurface has a smaller relative impact on the $S$-parameters. This is another distinction between a reverberant environment and an open one, where environmental losses only impact the signal magnitude through absorption.

We further show that our trained network can successfully determine the metasurface configuration from the measured scattering response in the cavity several days after the training data was collected. Measured $S_{21}$ responses with the same initial conditions inside a chaotic cavity will change over time, a phenomenon known as scattering fidelity decay \cite{SchaferExperimentalFidelity2005,taddeseSensorBasedExtending2009,taddeseSENSINGSMALLCHANGES}. Scattering fidelity decay is a property of wave chaotic system, and its sensitivity to boundary conditions and scattering environment.  This is in contrast to ray chaos and the sensitivity of bouncing ray trajectories to initial conditions in billiards. This decay means that any deep learning system that learns scattering responses inside a chaotic cavity will require periodic retraining. As discussed in Section \ref{sec:summary}, the fact that we are still able to determine the metasurface configuration accurately after several days means our technique is operationally useful, as it can function at a high level of accuracy for a long period of time before requiring retraining. Our approach is robust and highly accurate in determining metasurface commands from measured cavity $S_{21}$ spectra, providing an enabling capability for intelligent wavefront shaping applications. In addition, our method is general enough to operate in arbitrary complex scattering systems and does not requires a specifically engineered environment.

Our technique is achieved through the development and combination of four major aspects: 1) adaptive configuration of the metasurface unit cells by binning elements together to dynamically alter the relative size of the elements; 2) representation of the complex system $S$-parameters in a pseudo-2D ``image'' to promote extraction of features that are correlated over both local and global frequencies; 3) complex-valued deep learning layers to exploit both phase and amplitude information, accelerating training and improving the accuracy when applied to complex scattering environments; and 4) introduction of the Terrapin Module to parallelize the deep learning network, promoting sparse feature representation and improving training robustness.

\section{Experimental Configuration}
\label{sec:configuration}
\begin{figure}
\includegraphics[width = 8.6cm]{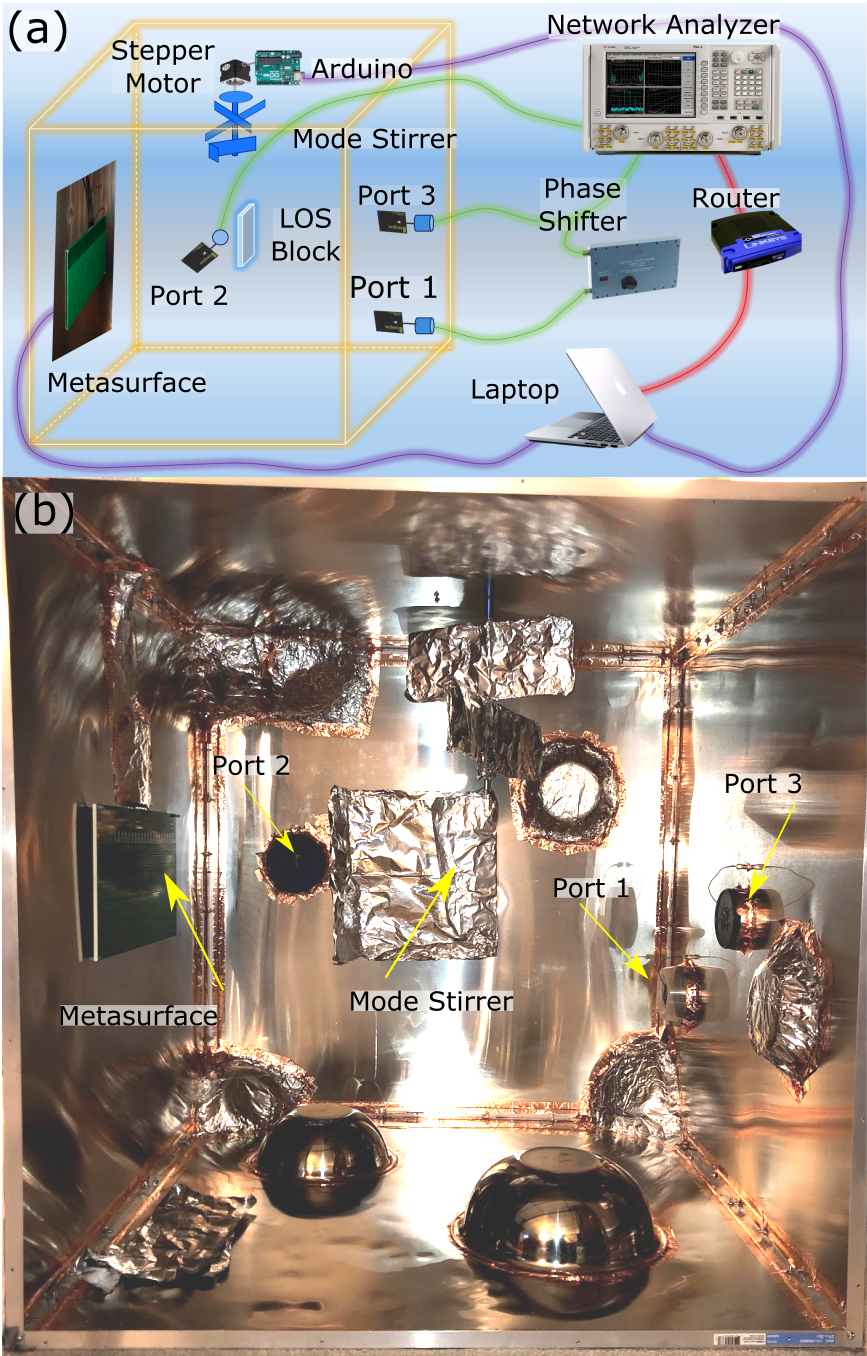}
\caption{\label{fig:cavity_configuration} \bf Cavity configuration. \normalfont \bf(a) \normalfont Experimental schematic of the cavity, showing the metasurface installed on the cavity walls, the locations of the 3 ports, the line-of-sight (LOS) block to prevent direct transmission between Port 2 and Ports 1 and 3, and the mode stirrer that is controlled by a stepper motor through an Arduino. Also shown are the network analyzer, phase shifter, control laptop and router. \bf (b) \normalfont Photograph of the interior of the cavity showing the components from the schematic as well as the irregular scatters that were installed on the cavity walls.}
\end{figure}

The complex, ray chaotic cavity used for experimentation is in the same configuration used in our previous work \cite{frazierWavefrontShaping2020}. It has a volume of $\sim$0.76 m$^3$ and includes 3 ports, with ports 1 and 3 used to inject signals and port 2 used for scoring. The cavity configuration and experimental schematic are shown in Fig. \ref{fig:cavity_configuration}. Each port is connected to an ultra wideband antenna (UWB) and the nominal measurement window is 3-4 GHz. An Arduino controlled mechanical mode stirrer is included to allow collection of an ensemble of cavity realizations. The experimental setup is controlled by a laptop, with an Agilent N5242A network analyzer used to measure cavity $S$-parameters. To reduce the cavity symmetry, irregular scattering objects were installed on the walls. Additional details on the cavity are provided in Appendix \ref{app:cavity}. 

Experiments were carried out over several months with the cavity placed in a heated and air-conditioned basement. While there were many commercial devices in the vicinity that emitted signals within the measurement window, we found the cavity to be well isolated. When not actively in use, such as in between experimental runs, the metasurface was powered off. 
 
The metasurface installed on an interior wall was fabricated by the Johns Hopkins University Applied Physics Laboratory. It is designed to operate in the frequency range of 3-4 GHz, and contains 240 binary meta-atoms (LC resonators) arranged in a rectangular grid of $10 \times 24$ elements. Each element has a characteristic length of $\sim \lambda/6$ and is switched by a GaAs transistor amplifier to 1 of 2 states (0 or 1), changing the phase of the reflection coefficient by $\sim 180^{\circ}$ \cite{schmidSbandGaAsFET2020}. The metasurface covers a small region of the interior surface area of the cavity, $1.5\%$, and intercepts only a limited number of rays. 

The goal of our deep learning network is to enable wavefront reconstruction inside a complex cavity. The network will accept a given $S_{21}$ spectra from 3-4 GHz and accurately determine the metasurface commands necessary to closely realize that specific scattering response. The relatively small size of the metasurface and its unit cells leads to high correlation between system scattering responses with minor changes in metasurface commands, which means the inverse problem is ill-posed. Deng et al. recently introduced a neural-adjoint approach for solving the ill-posed inverse problem of designing unit cell geometries to match specified absorption spectra \cite{DengNeuralAdjoint2021}. In this case, a fully connected deep learning network was used to model the forward problem, acting as a Green's function to predict the spectrum from a given design. The inverse problem was then solved iteratively, driving the design along an estimated gradient towards the optimal solution. As discussed in Appendix \ref{app:wavefront_reconstruction_control}, gradient methods work best for a continuous or near continuous solution space rather than a binary one such as ours; however, the adjoint method from \cite{DengNeuralAdjoint2021} can be adapted into a reinforcement learning approach \cite{bertsekasReinforcementLearningOptimal2019}. In addition, inside a chaotic reverberating environment, the spectra will have more structure, resulting in higher frequency oscillations or local features that must also be learned. Therefore, we require a different deep learning approach.

\section{Deep Learning Network Design}
\label{sec:deep_learning_network_design}
The metasurface has $2^{240}$ possible combinations of commands. To reduce the dimensionality, we introduced the concept of binning neighboring unit cells as discussed in Appendix \ref{app:metasurface_binning}. Groupings consisting of $2\times2$, $3\times3$, and $5\times4$ elements were chosen. Binning is an important capability that allows us to adapt the size of an effective element to the underlying scattering system. This is one of the major contributions of our work.

The primary limitation of our approach is that we are not guaranteed to be able to generate any arbitrary $S_{21}$ response, as a configuration of the metasurface that produces that response does not necessarily exist. The small size of the metasurface relative to the cavity limits its ability to interact with all possible ray trajectories, emphasizing the importance of a binning capability to adapt the effective pixel size to the environment. This limitation is therefore a function of the system configuration, and not the deep learning network. The small relative size of the metasurface does represent a realistic configuration for practical smart radio environments, however.

\begin{figure*}
\includegraphics[width = 17.2cm]{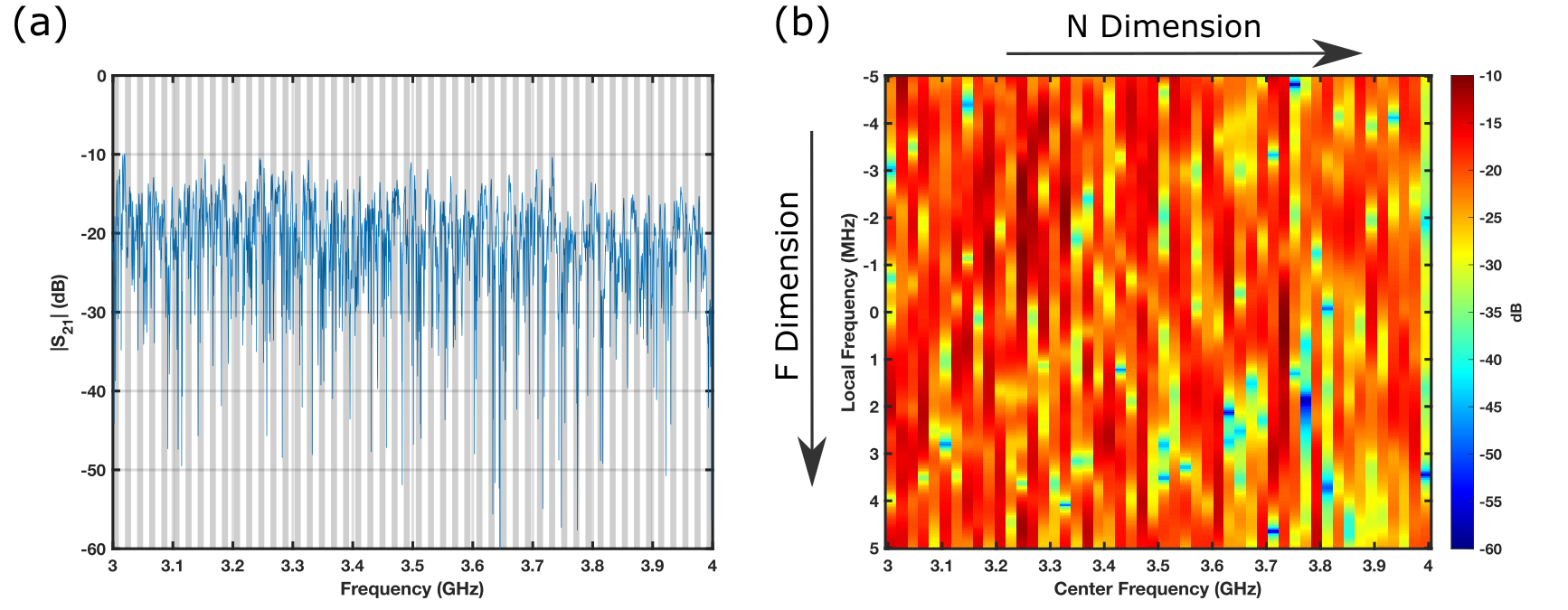}
\caption{\label{fig:dataPreparation} \bf Data preparation example. \normalfont The deep learning networks use complex amplitudes, however, only the magnitude is shown here for illustrative purposes. \bf (a) \normalfont Raw $|S_{21}|$ data vs. frequency showing 50 local windows highlighted in gray. The actual data preparation uses 100 local windows, only 50 are shown here for clarity. The data was collected over a 1 GHz measurement window with 32,001 points and each local window (highlighted in gray) has a bandwidth of 10 MHz or 321 points. \bf (b) \normalfont Extracted $|S_{21}|$ data (in log scale) in a pseudo-2D format as a $321\times 50$ pixel ``image". The data is represented as center frequency (over the full 1 GHz measurement window or $N$ dimension) vs. local frequency (over the 10 MHz local window or $F$ dimension). The deep learning network will use 1D convolutions to extract features in the 10 MHz local frequency windows ($y$-axis) and use the relationships between convolutional filters to capture global correlations over the full 1 GHz measurement window ($x$-axis).}
\end{figure*}

An important step for deep learning is preparation of the measured data. The goal here is to represent the data in a basis set that can be ingested by the neural network architecture. The raw data consists of $M$ sets of complex two-port $S$-parameter values, each containing 32,001 points measured over a 3-4 GHz window. We are interested in the relationship between metasurface commands and transmission between the ports, so we select $S_{21}$ as the primary variable of interest. The measured data contains local and global correlations, both of which must be captured by the deep learning network. We can exploit the local correlations with 1D convolutional neural network (CNN) layers, but would like the individual windows to cover a smaller bandwidth. Our previous work showed diminishing returns for optimization over bandwidths greater than 10 MHz \cite{frazierWavefrontShaping2020}, so 10 MHz provides a reasonable limit for the local window size. We therefore extract the complex $S_{21}$ in 10 MHz frequency windows at 100 distributed center frequencies to provide 100 feature vectors containing 321 points each. A representative data set is shown in Fig. \ref{fig:dataPreparation} (a), with only 50 feature vectors used for illustration. The data are organized into a 3D structure of $M$ sets of data $\times$ $F$ local frequencies $\times$ $N$ features, or $10,000 \times 321 \times 100$ for the $2\times2$ binning configuration. Each data set takes on a pseudo-2D format with a $321\times 100$ pixel ``image" as shown in Fig. \ref{fig:dataPreparation} (b). Local features in the 10 MHz frequency windows (over the $F$ dimension) will be extracted by 1D CNN layers and global features (over the $N$ dimension) will be extracted by the overall deep learning network architecture. The overall architecture then acts as a dense or fully connected layer from the perspective of the global features. The pseudo-2D format and its ability to capture both long-range and short-range correlations in frequency provides the second major aspect of our approach. Additional details on data collection are provided in Appendix \ref{app:data_prep}. 

The output values of the deep learning network (equal in number to the number of binned metasurface elements) are floating point numbers rather than binary numbers and can be interpreted as the probability that a given element in the metasurface is active (set to 1). The determined commands are then found by rounding the outputs to either a 1 or a 0. Inspection of the raw (unrounded) outputs allows us to assess how correct the deep learning network was, or how confident the network was in the result.  A discussion of the different types of neural network layers used is provided in Appendices \ref{app:deep_learning} and \ref{app:1dConvolution}.

As described in Appendix \ref{app:network_training}, the training data sets are randomly shuffled and then split into 75\% training data and 25\% validation data. At each step (epoch), validation of the trained model is performed by testing the model with a new set of data not present in the training set.

Complex-valued multiplicative layers have been used to invert propagation through multi-mode fibers \cite{moranDeepComplex2018,caramazzaTransmissionNaturalScene2019}, but , have not previously been used for wavefront reconstruction. Unfortunately, as discussed in Appendix \ref{app:complex_layers}, there are no officially supported complex-valued modules in any of the major deep learning frameworks. Multiplicative layers are straightforward to implement, but more complicated modules, such as convolutional layers, are not. For the research described here, we leveraged the open source complexPyTorch library \cite{popoffComplexPyTorch2019} as the basis for our complex-valued network layers.

The added complexity resulting from placing the metasurface inside a chaotic cavity requires a correspondingly complicated deep learning network to extract the relevant features. Rather than only designing progressively more intricate network topologies, we can also introduce complex-valued layers \cite{trabelsiDeepComplexNetworks2018}, which serve as the third major aspect of our approach. The wave scattering phenomenon is fundamentally complex-valued, so using complex-valued layers allows the network to exploit both phase and amplitude information and better match the underlying physical system. 

\begin{figure*}
\includegraphics[width = 17.2cm]{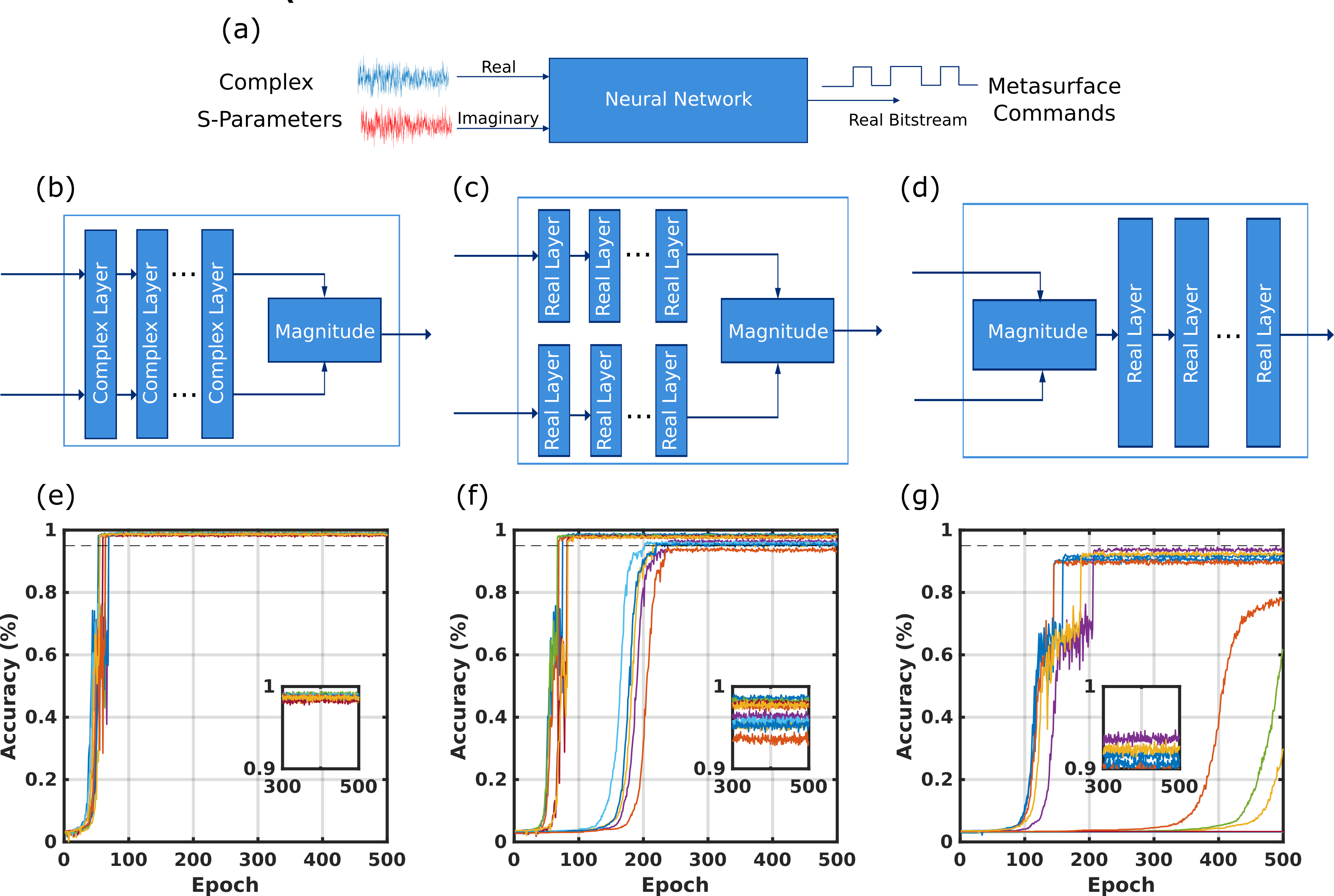}
\caption{\label{fig:validationArchitecture} \bf Training performance for different architectures. (a) \normalfont Generic network architecture. \normalfont The real and imaginary components of the complex $S$-parameters are ingested by the network, with a real-valued bitstream provided at the output for metasurface commands. \bf (b) \normalfont Schematic architecture for the complex-valued network showing that the magnitude operation only occurs at the end of the network. \bf (c) \normalfont Schematic architecture for the hybrid network showing that the real and imaginary components are processed independently and combined at the end. \bf (d) \normalfont Schematic architecture for the real-valued network showing that the magnitude operation occurs at the beginning of the network. \bf (e) \normalfont through \bf (g) \normalfont Evolution of the accuracy over the validation data set for the various architectures with a mean accuracy of 99.2\% for the complex valued network \bf (e)\normalfont, 97.4\% for the hybrid network \bf (f)\normalfont, and 91.4\% for the real-valued network \bf (g) \normalfont cases that successfully trained. The dashed line shows 95\% accuracy, and the insets show closeups of the accuracy evolution from epochs 300-500 in the range of 90\% to 100\%. This figure demonstrates that the complex-valued network achieves better overall accuracy with more stable training performance than the other architectures.}
\end{figure*}

To demonstrate the impact of utilizing complex-valued network layers, we performed an experiment that compared 3 different architectures: 1) a complex-valued network that processes the complex-valued $S$-parameters and converts the result to a magnitude at the end of the network (Fig \ref{fig:validationArchitecture}b); 2) a hybrid network that processes the real and imaginary components of the $S$-parameters independently and then combines them through a root-sum-square operation at the end of the network (Fig \ref{fig:validationArchitecture}c); and 3) a real-valued network that operates on the magnitude of the $S$-parameters (Fig \ref{fig:validationArchitecture}d).

Ten separate training runs were performed with each architecture on the same set of input data. Panels e) through g) of Figure \ref{fig:validationArchitecture} show the evolution of the accuracy of network over the validation data set and demonstrate that the complex-valued network layers are ultimately able to achieve a higher accuracy than the hybrid or real-valued network layers and exhibit more stable training behavior. Training instability manifests as wide variations in the accuracy response, resulting in certain runs not effectively learning until much later (several cases for the real-valued network are still learning after 500 epochs). It indicates that the hybrid and real-valued networks are much more sensitive to initial weights or ordering of the data sets after shuffling than the complex-valued networks.

As described in Appendices \ref{app:network_architecture} and \ref{app:bin5x4}, the $5\times 4$ binning case performed extremely well using a straightforward sequential CNN architecture. After training, we were able to accurately realize 100\% of the target responses over both the training and validation sets. Unfortunately, the purely sequential architecture of the network did not work as well for the $3\times 3$ binning configuration (see Appendix \ref{app:3x3}). The increased complexity implies that we need a more complex network, so we turned to approaches successfully used in modern image classification, specifically inception modules \cite{szegedyGoingDeeperConvolutions2015, szegedyRethinkingInceptionArchitecture2016}. As discussed in Appendix \ref{app:receptive_field}, we modified the general architecture to perform 1D convolutions over the 10 MHz local frequency windows. The 1D convolutional filters then extract local features over the 10 MHz windows, while the relationship between the filters acts as a dense or fully connected layer, extracting global features over the full 1 GHz measurement window. The final version, which we refer to as a ``Terrapin Module'', is shown schematically in Fig. \ref{fig:modified_inception_module} and provides the fourth and final major technical contribution of our approach.

With a deep learning network containing 4 Terrapin Modules in series, we were able to get excellent performance for the $3\times 3$ binning configuration with only 4,000 sets of training data, as discussed in Appendix \ref{app:3x3}. The $2\times2$ configuration required 10,000 sets of data for a similar level of performance (see Appendix \ref{app:2x2}). The smaller effective elements in this configuration produce responses with a larger degree of correlation. Thus, more data is required for the network to learn and distinguish the more subtle relationships between metasurface command configurations and scattering matrix responses, $S_{21}(f)$.  

Training is performed in a parallelized fashion over all the training data at once, e.g., for our computational resources,  taking $\sim$ 4 hours to collect a sufficient set of data and train the deep learning network. Testing, however, is performed on single shot measurements and takes less than 1 second to measure the $S_{21}$ response and make a determination of the metasurface commands, enabling real-time operation. In contrast, the iterative approach in our earlier work \cite{frazierWavefrontShaping2020} required $\sim$300 measurements to converge to a desired configuration, taking $\sim$10 minutes to reach the answer for each configuration. Online iterative optimization does not require training, but may produce undesirable configurations due to the randomly applied perturbations. When time is available for offline training, the deep learning approach is preferred.

\section{\label{sec:online} Results}
The primary objective of this work is to demonstrate that deep wavefront shaping is a viable technique for wavefront reconstruction inside complex scattering environments, enabling intelligent wavefront shaping in a chaotic cavity. In this section, we shown how our deep learning approach accurately determines metasurface commands from measured cavity scattering parameters.

Training results for the $5\times4$ and $3\times3$ binning configurations are provided in Appendices \ref{app:bin5x4} and \ref{app:3x3}, while training results for the $2\times 2$ binning configuration are provided in Appendix \ref{app:2x2} and shown in Fig. \ref{fig:performance_2x2_binning}. The training data consists of 10,000 random realizations of metasurface commands, representing an extremely small fraction of the 1.2 $\times 10^{18}$ possible configurations. The data was split into 7500 sets for training and 2500 sets for validation to ensure the validation process is unbiased. The evolution of the loss function is shown in Fig. \ref{fig:performance_2x2_binning} (a) and the evolution of the accuracy is shown in Fig. \ref{fig:performance_2x2_binning} (b). The loss function was chosen as the mean absolute error, or the $\mathcal{L}_1$ norm, between determined and actual metasurface commands. Accuracy is the fraction of sets that was determined without error and provides a more conservative estimate of performance. The variation around Epoch 45 is due to choosing an aggressive initial learning rate and the impact of reducing the  learning rate on a plateau can be seen at Epoch 55. These panels demonstrate that we were able to obtain high accuracy and a small loss function for both the training and validation data sets.

The trained model did not have perfect accuracy but was able to determine 2,443 out of 2,500 sets in the validation data without error for an accuracy of 97.7\%. One set had 2 errors and 56 sets had a single error, as shown in Fig. \ref{fig:performance_2x2_binning} (c). A comparison of the determined and true commands is shown in Fig. \ref{fig:performance_2x2_binning} (d), (e), (g), and (h). These panels show that for the worst case set with two errors, the network was not highly confident in the results as the erroneous determined command probabilities were 0.41 and 0.73.  Finally, example scattering responses are shown in Fig. \ref{fig:performance_2x2_binning} (f) and (i), which demonstrate both the complexity of the $S_{21}$ responses and the fact that the difference between the measured and predicted responses are generally at least 20 dB lower than the signal magnitudes themselves.  

\begin{figure*}
\includegraphics[width = 17.2cm]{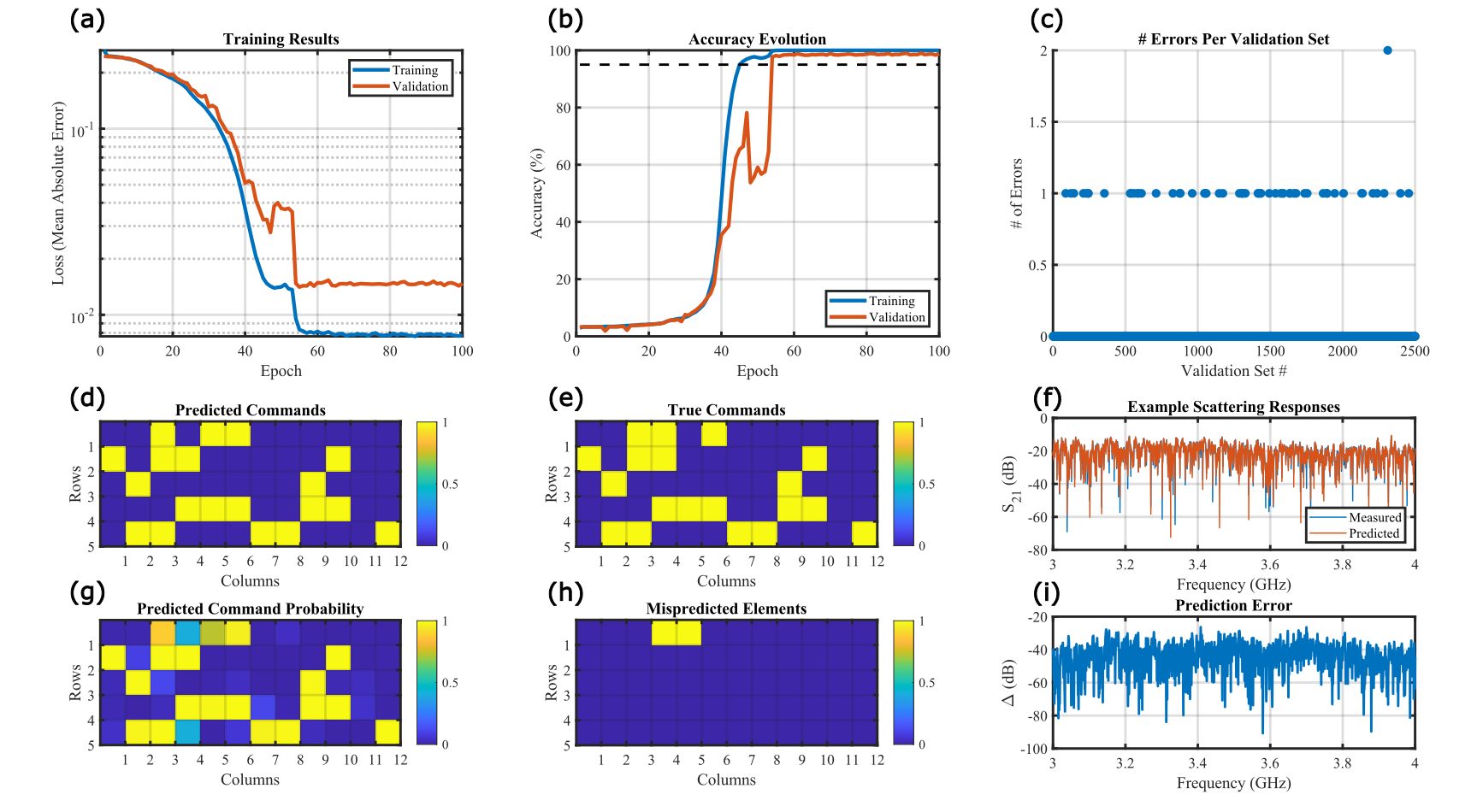}
\caption{\label{fig:performance_2x2_binning} \bf Deep learning performance with complex-valued layers for 2x2 binning. \normalfont \bf (a) \normalfont Evolution of the loss function for the training and validation sets over 100 epochs. The learning rate is reduced at Epoch 55, inducing an additional reduction in the loss function. The initial learning rate was aggressive, resulting in a large variation in the validation loss function between Epochs 45-50. \bf (b) \normalfont Evolution of the accuracy for the training and validation sets over 100 epochs, the dashed black line identifies 95\% accuracy. \bf (c) \normalfont Number of errors over the validation set for the trained model, showing a total of 58 errors and 97.7\% accuracy. The maximum number of errors in a single set was 2 (out of 60 elements), which occurred once. \bf (d) \normalfont Determined commands for validation set \#2311 showing the output from the deep learning network. \bf (e) \normalfont True commands for validation set \#2311 showing what was actually applied to the metasurface. \bf (f) \normalfont Example scattering responses for online validation showing the measured and predicted $S_{21}$responses. \bf (g) \normalfont Determined command probability for validation set \#2311, showing the raw outputs from the deep learning network prior to rounding. This panel shows that the 2 elements determined incorrectly have command probabilities of 0.41 and 0.73, meaning the network was not highly confident in the result. \bf (h) \normalfont Errors, or incorrectly determined commands for validation set \#2311, showing the 2 elements that were determined incorrectly. \bf (i) \normalfont Prediction errors or difference between the measured and predicted $S_{21}$ responses. }
\end{figure*}

\begin{figure*}
\includegraphics[width = 17.2cm]{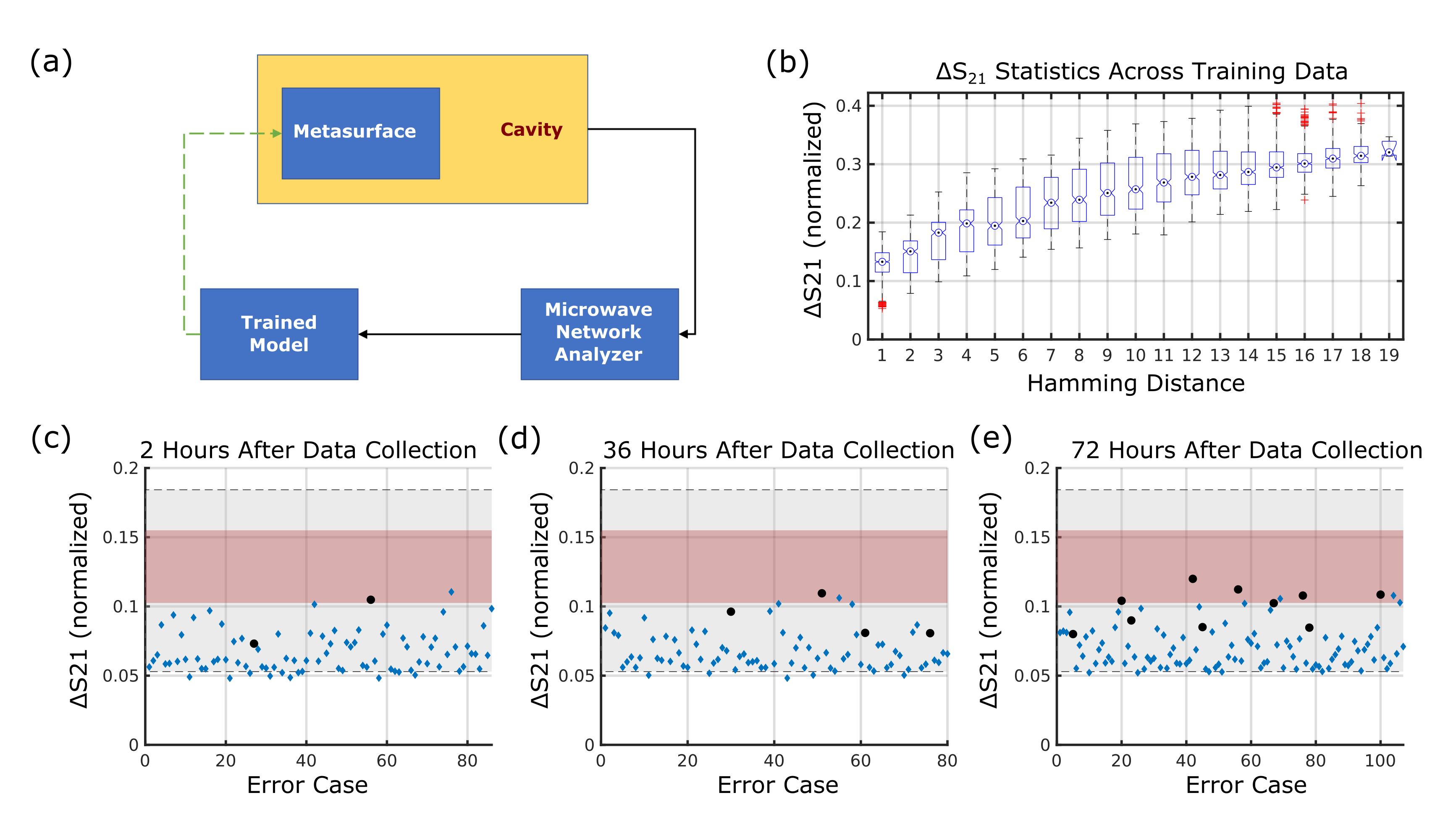}
\caption{\label{fig:online_validation} \bf On-line performance verification. \normalfont \bf (a) \normalfont Closed loop validation configuration. Commands were applied to the metasurface inside the cavity, the corresponding $S_{21}$ response was measured on the PNA, and the results were passed through the trained deep learning network. Errors for the trained model were then measured to determine the difference in $S_{21}$, $\Delta S_{21}$, between the two command sets. \bf (b) \normalfont $\Delta S_{21}$ statistics for the minimum Hamming distance across the 10,000 sets from the training data. Whisker plots are given for the smallest Hamming distance for each case, and show the mean value, 25$^{\text{th}}$ and 75$^{\text{th}}$ percentiles, and maximum and minimum values. \bf (c) \normalfont through \bf (e) \normalfont $\Delta S_{21}$ for online validation sets taken a specified time after the training data was collected. The shaded regions show the extent of the single element Hamming distance whisker box plot from panel (b). The grey region shows the full range from maximum to minimum, and the red region shows the 25$^{\text{th}}$ and 75$^{\text{th}}$ percentiles. The blue diamonds indicate cases with a single error, while the black circles indicate cases with 2 errors. These panels show that the $\Delta S_{21}$ for errors is very small, and in the lower region of the statistics covered by observed cases with single element Hamming distances. \bf (c) \normalfont Validation 2 hours after collecting training data, 2000 sets of commands were tested with 86 errors for an accuracy of 95.7\%.  \bf (d) \normalfont Validation 36 hours after collecting training data, 2000 sets of commands were tested with 80 errors for an accuracy of 96\%. \bf (e) \normalfont Validation 72 hours after collecting training data, 2000 sets of commands were tested with 107 errors for an accuracy of 94.7\%. }
\end{figure*}

To further validate the trained deep learning network, we adopted the on-line, closed loop configuration as shown in Fig. \ref{fig:online_validation} (a). Commands were applied to the metasurface and the $S_{21}$ response was measured and then passed through the trained deep learning model to verify accuracy. This provides a 3rd set of data that was not seen during training (or the initial validation). When the deep learning determined commands had errors, the determined commands were applied to the metasurface so that the difference in $S_{21}$ responses could be computed. We define the difference, $\Delta S_{21}$, between two measured $S_{21}$ responses, $S_{21}^a$ and $S_{21}^b$ through the $\mathcal{L}_2$ (Euclidean) norm, $||S_{21}(f)||_2 = \sqrt{\sum_f |S_{21}(f)|^2}$. The summation is taken over the full measured frequency range (3-4 GHz) and $\Delta S_{21}$ is defined as

\begin{equation}
    \Delta S_{21} = 2\frac{||S_{21}^a(f) - S_{21}^b(f) ||_2}{||S_{21}^0(f)||_2 + ||S_{21}^1 (f)||_2}
\end{equation}

The normalization factor here is determined by the average of the $\mathcal{L}_2$ norms of the active commands (all 1s), $S_{21}^1$, and the inactive commands (all 0s), $S_{21}^0$. To understand how $\Delta S_{21}$ depends on the difference between commands, we first identified the minimum Hamming distance for each of the 10,000 sets in the training data. The Hamming distance is simply the number of elements that are different between 2 sets of commands. It is a useful metric for comparing command sets, but does not include scaling or correlation based on position; in some cases, the impact of toggling an element in the center may be significantly different than the impact of toggling an element on the edge of the metasurface.  

The smallest Hamming distance between the training data sets ranged from a single element to 19 elements (out of 60). A series of whisker box plots showing $\Delta S_{21}$ for the various Hamming distances is shown in Fig. \ref{fig:online_validation} (b). The general trend shows an increase in $\Delta S_{21}$ with an increase in the Hamming distance. While the relationship is nonlinear, the dynamic range in $\Delta S_{21}$ for Hamming distances up to 1/3 of the total number of elements is large, approximately an order of magnitude.

Validation of the deep learning network in the configuration shown in Fig. \ref{fig:online_validation} (a) was performed periodically after collecting the training data and the results are shown in Fig. \ref{fig:online_validation} (c) at 2 hours, (d) at 36 hours (1.5 days), and (e) at 72 hours (3 days), with the metasurface powered off between each validation test. Over time the scattering environment is expected to ``age'' and systematic changes to the scattering environment will occur. The blue diamonds show cases where there was a single error, and the black dots show cases where there were 2 errors. Each on-line validation experiment showed $\sim 95$\% accuracy and the resulting $\Delta S_{21}$ for errors was small compared to the observed statistical extent of $\Delta S_{21}$ for single element Hamming distances. This suggests that even when the deep learning network is unable to determine the commands completely accurately, the resulting difference in $S_{21}$ is very small. As shown in Appendix \ref{app:scattering_fidelity}, the accuracy was still $>$85\% after 120 hours (5 days), but dropped to $\sim$65\% after 9 days.

The number of errors and number of cases with more than one error increases with time, showing the ``aging'' effect of the cavity, which can be quantified through the concept of scattering fidelity. Scattering fidelity is the normalized correlation as a function of time between two cavity responses with the same initial conditions \cite{taddeseQuantifying2013}. Because a chaotic cavity is sensitive to small changes in the boundary conditions, such as volume perturbations, the scattering fidelity will decay over time \cite{SchaferExperimentalFidelity2005, taddeseSensorBasedExtending2009,taddeseSENSINGSMALLCHANGES}. Loss of scattering fidelity means that the accuracy of any trained deep learning network has a finite lifetime, so we must periodically retrain the network on new training data to maintain accuracy. In our case, we have demonstrated that the deep learning network can determine metasurface commands with high accuracy ($>$ 95\%) for at least 72 hours (3 days) after the initial training data collection, and with reasonable accuracy ($>$ 85\%) up to 120 hours (5 days) after the initial training data collection. Large variations in environmental conditions, such as temperature or humidity, will introduce larger perturbations and more rapidly degrade the scattering fidelity.

An additional set of experiments showing the performance of the deep learning network with different cavity reverberation times is provided in Appendix \ref{app:reverberation_time} and Fig. \ref{fig:reverb}.

\section{\label{sec:summary}Summary and Discussion}
In this paper, we demonstrated the use of a deep learning network for wavefront reconstruction, enabling intelligent wavefront shaping in complex reverberant environments. Major aspects include complex-valued deep learning layers that exploit both phase and amplitude information, binning of the metasurface elements, a pseudo-2D data format that allows features to be extracted over both narrow and wide bandwidths, and a Terrapin Module that enhances the receptive field, providing width and depth to the network. While our configuration leverages a binary (1-bit) tunable metasurface, it can be adapted for a continuous device, such as a varactor, by replacing the output sigmoid activation function with a rectified linear unit activation function and then discretizing the signal to the desired resolution.

One of the primary limitations of traditional deep learning is the amount of data required to train the networks. This is especially concerning in light of the fact that loss of scattering fidelity requires periodically collecting new training data. We have demonstrated the ability to train highly accurate networks with a limited amount of training data, requiring far fewer sets than the number of possible combinations of commands. We have also demonstrated that the accuracy can be maintained for a period of at least several days, and is successful with varying amounts of loss in the cavity. This indicates that successful training on a reduced amount of data is possible, provided it is sufficiently diverse. Diversity in both the metasurface commands and the measured responses is then a key aspect in setting up any potential autonomous system.

Several concerns must be addressed to enable practical fielded hardware systems. First, the sensing component must be reduced in cost and size. The availability of software defined radio (SDR) architectures presents an ideal path here, with many inexpensive platforms readily available. Compact devices such as the bladeRF \cite{nuandBladeRF} can replace the bulky network analyzer. SDRs have limited instantaneous bandwidth, typically 10-20 MHz, so modifications would be required to the pseudo-2D data representation. Second, processing large deep learning models on power hungry GPUs may exceed the allowable footprint in terms of both cost and power consumption. Deep learning models can be compressed by pruning and quantization \cite{hanDeepCompressionCompressing2016}, and the explosion of edge intelligence for connected devices in the Internet of Things is leading to more efficient embedded deep learning systems. An example is the Jetson series of embedded GPUs from NVIDIA; the currently available TX2 series can provide up to 1.26 trillion floating point operations per second on a 256-core GPU while consuming only 10-20 W of power \cite{nvidiaJetson}.

In closing, we have shown that deep learning enabled wavefront reconstruction provides an important step towards realizing intelligent reconfigurable metasurfaces for smart radio environments. Potential applications in the domain of electromagnetics include wireless power transfer, protection of sensitive electronic components, optimization of wireless networks, micromanipulation of objects, and nonlinear time reversal. Our technique is applicable to general wave chaotic scattering systems and is not strictly limited to electromagnetic waves. Adopting this technique to control the system scattering response with metasurfaces that interact with seismic waves \cite{palermoControl2018,heSeismic2020} or quantum waves \cite{bekensteinQuantum2020} will unlock many innovative applications for wave chaotic systems.

\section*{\label{sec:acknowledgements}Acknowledgements}
We thank Zerotti Woods of JHU/APL for supporting many fruitful discussions on applied deep learning. Funding for this work was provided through AFOSR COE Grant FA9550-15-1-0171 and ONR Grant N000141912481. 


%

\appendix

\renewcommand{\theequation}{A\arabic{equation}}
\renewcommand{\thefigure}{A\arabic{figure}}
\setcounter{figure}{0}   
\renewcommand{\thesection}{A\arabic{section}}

\section{Wavefront Reconstruction and Control}
\label{app:wavefront_reconstruction_control}
Wavefront control has a rich history and has been well studied from the perspective of adaptive optics. Conceived by Horace Babcock in 1953 \cite{babcockAO1953}, and realized in the 1970's \cite{duffnerAORevolution2009}, adaptive optics provides a method of correcting wavefront aberrations induced by propagation through random media. It has been successfully used in a diverse array of applications including astronomical imaging \cite{hardyAOTelescopes1998}, biomedical imaging \cite{boothAOMicroscopy2007}, high energy laser propagation \cite{HoggeAOHEL1978}, free space optical communications \cite{whiteAOComms2004}, quantum networking with satellites \cite{gruneisenQuantum2020}, and laser processing of  materials \cite{salterAOProcessing2019}.

Adaptive optics systems typically employ reflective deformable mirrors that provide mechanical phase compensation \cite{FreemanDM1982,ealeyDM1989}. Conventional deformable mirrors are built with ferroelectric actuators \cite{ealeyActuators1992} with as small as 5 mm spacing, though monolithic deformable mirrors manufactured from a block of lead magnesium niobate (PMN) material can have 1 mm spacing between actuators \cite{wirthDMTech2013}. Micro-electrical-mechanical-system (MEMS) mirrors using electrostatic actuation have made great strides over the past decade \cite{horensteinMEMS1999,perreaultMEMS2002} and are very competitive with conventional deformable mirrors, particularly where a large actuator density is desired. Refractive liquid crystal devices have also been proposed and developed \cite{casasentSLM1977,espositoLC1993}, but tend to be slow and are uncommon outside of microscopy \cite{colliniAdaptiveOpticsMicrospectrometer2019}, or laser processing \cite{salterAOProcessing2019}.

The availability of inexpensive reconfigurable metasurfaces has driven research into a field known as wavefront shaping \cite{vellekoopFocusingCoherentLight2007,vellekoopUniversal2008,moskControl2012,kainaShapingComplexMicrowave2015,vekslerMultipleWavefrontShaping2015,delhougneShapingMicrowaveFields2017}. While there is not a strict convention or definition, adaptive optics is generally associated with controlling distorted wavefronts for a single propagation path while wavefront shaping is generally associated with controlling (or shaping) a combination of multiple scattered wavefronts. We will adopt this convention here and refer to adaptive optics and wavefront shaping in general as wavefront control. While many applications are built around scattering systems possessing time reversal symmetry (TRS), the presence of TRS is not a requirement for all wavefront control systems. 

Conventional adaptive optics systems measure the wavefront directly and use an operator, called the reconstructor, to solve the inverse problem between measurements and control signals. Wavefront reconstruction is at the heart of any wavefront control system. The process is generally indirect, as the reconstructor evaluates the wavefront in the basis of command signals, rather than explicitly in phase. Wavefront reconstruction is a specialized area of system identification \cite{ljungSystemIdentification1997}, and relies heavily on methods for solving inverse problems.

For a linear system, or one that can be linearized, the standard reconstructor, $\mathbf{R}$, is a regularized optimal Wiener filter given as \cite{tysonFrazierfieldGuide2012}

\begin{equation}
\label{eq:reconstructor}
   \mathbf{R} = \left[\mathbf{B}^T\mathbf{C}_n^{-1}\mathbf{B} + \mathbf{W} + \mathbf{B}^T\mathbf{C}_{\varphi}^{-1}\mathbf{B} \right]^{-1}\mathbf{B}^T\mathbf{C}_n^{-1}
\end{equation}

\noindent Here, $\mathbf{C}_n$ is the measurement noise covariance matrix, $\mathbf{C}_n = \left<\mathbf{n}^T \mathbf{n} \right>$, $\mathbf{C}_{\varphi}$ is the covariance matrix of the environmental disturbance, $\mathbf{C}_{\varphi} = \left<\mathbf{\varphi}^T \mathbf{\varphi} \right>$, $\mathbf{W}$ is a weighting/regularization matrix, and $\mathbf{B}$ is a system configuration (geometry) matrix that relates control signals to wavefront measurements. Both $\mathbf{C}_n$ and $\mathbf{C}_{\varphi}$ are defined in sensor space, while $\mathbf{W}$ is defined in command space. 

The inverse problem is often ill-posed due to singularities or the presence of highly correlated responses with different commands \cite{neubauerIllPosed2010}, which effectively means we do not have enough information to solve the problem. We can add the necessary information through a process known as regularization. The weighting matrix, $\mathbf{W}$, in Eq. \ref{eq:reconstructor} implicitly provides Tikhonov regularization, which acts as a spectral filter on the singular values \cite{vogelInverse2002,hansenDeblurring2006}. $\mathbf{W}$ can be the identity matrix to raise all the singular values, or a projection matrix to suppress specific modes as they may either induce singularities or expend control energy in ways we wish to avoid. Using the inverse of the environmental covariance matrix, $\mathbf{C}_{\varphi}^{-1}$, preconditions the solution towards expected spatial modes with the appropriate spatial statistics.

For nonlinear systems, we can apply iterative methods to handle the reconstruction process. These are typically Krylov subspace methods such as the conjugate gradient method \cite{vogelInverse2002,HuttererNonlinear2018,HuttererIterative2019}. Regularization can be applied through Landweber iteration, where the gradient is allowed to decay with a relaxation parameter \cite{vogelInverse2002,HuttererNonlinear2018}. In this case, the iterations are performed ``offline'', meaning that each iteration is evaluated numerically on the measurements. The convergence rate is therefore only limited by the computational power we can throw at the problem.

Wavefront reconstruction is viable for some metasurface applications, but depends on being able to solve the inverse problem. Complex environments include uncertainties in determining the system configuration ($\mathbf{B}$ matrix), and multiple reflection paths create intricate interference patterns at the antennas, producing chaotic fluctuations \cite{hemmadyUniversalStatisticsScattering2005,hemmadyUniversalPropertiesTwoport2006}. In addition, short orbits that manifest as persistent features in the ensemble \cite{hartEffectShortRay2009} are not removed. This leads to a complicated relationship between metasurface commands and cavity scattering parameters. For a metasurface that is small relative to the cavity, the effective strength of the metasurface commands on the cavity scattering parameters is reduced. This results in high correlation between measurements taken with different sets of commands and creates problems for uniqueness, as many potential solutions are extremely similar. In addition, the scattering process is linear, but the relationship between metasurface commands and measured scattering parameters is not necessarily so, particularly for measurements in the temporal domain. In these extreme scattering environments, we are limited to partial information and may not be able to define the system, let alone determine the inverse. This leads to model-free control approaches that do not require knowledge of the system configuration, and bypass an explicit wavefront reconstruction step altogether.

Early metasurface control approaches used brute force trial and error, toggling every element or combination of elements \cite{vellekoopPhase2008, dorrerDirectBinarySearch2018}. This guarantees that a global minimum is reached, but becomes infeasible with large numbers of elements. The simplest practical approach is an extension of iterative multidither techniques in adaptive optics \cite{OMearaMultidither1977}, where, at the $k^{\text{th}}$ iteration, the algorithm updates a trial command vector, $\mathbf{a}^*$, with a small perturbation, $\delta \mathbf{a}$, so that

\begin{equation}
\label{eq:dither}
    \mathbf{a}^*_{k+1} = \mathbf{a}_k + \delta \mathbf{a}_k
\end{equation}

The impact on performance is evaluated through a metric or cost function, $J$, that is positive and real-valued, and dependent on both the command vector and the environment, $\mathcal{E}$. If the cost function, $J\left(\mathbf{a}^*_{k+1},\mathcal{E}\right)$, is improved, the trial command vector becomes the new command vector, $\mathbf{a}_{k+1} = \mathbf{a}^*_{k+1}$. Otherwise, the trial command vector is rejected and a new trial command vector is generated. The iterative process continues until either a specified number of iterations, $T$, are performed without improving the metric, or the cost function reaches a pre-determined value, at which point we claim convergence. While simple to implement and not reliant on knowledge of the system configuration, the dithering approach is by no means optimal.

Gradient based approaches have proven extremely successful for general stochastic optimization problems \cite{spallOptimizationBook2003}. In a stochastic gradient descent (SGD) optimization, the descent is performed by taking steps along the gradient of the cost function with respect to the element command vector. The step size, $\gamma$, which may or may not depend on the iteration, determines how quickly the algorithm descends, and is sometimes referred to as a ``learning rate''. Tuning the step size is an important aspect of SGD methods. If $\gamma$ is too small, the algorithm will take a long time to converge and may not be able to escape a local minimum. On the other hand, if $\gamma$ is too large, the algorithm may become unstable. The basic SGD is implemented as

\begin{equation}
\label{eq:sgd}
\mathbf{a}_{k+1} = \mathbf{a}_k - \gamma_k \nabla_{\mathbf{a}_k} J\left(\mathbf{a}_k,\mathcal{E}\right)
\end{equation}

In most cases, it is not possible to evaluate the gradient, $\nabla_{\mathbf{a}} J$, directly, so it must be approximated. The general approach is to apply a small perturbation to the current command vector and estimate the gradient from a one-sided or two-sided finite difference. The perturbation is applied to all elements of the command vector simultaneously (in parallel) to increase the convergence rate.

A specialty of wavefront control, known as wavefront sensorless, or target-in-the-loop adaptive optics \cite{VorontsovTargetInTheLoop2005}, leverages stochastic optimization in a sensor agnostic manner, indirectly evaluating the wavefront through a scalar cost function, $J$. In target-in-the-loop approaches, the iterations are performed ``online'', meaning that each iteration requires applying commands and measuring the result. The convergence rate is therefore limited by the sampling rate of the system. 

Target-in-the-loop methods are not as easily analyzed through modern multivariable control theory as conventional methods, but they are highly applicable to the problem of controlling metasurfaces in complex scattering environments. In particular, stochastic parallel gradient descent (SPGD) \cite{VorontsovSPGD97,VorontsovSPGD98} has enjoyed great success in target-in-the-loop adaptive optics systems. For SPGD, the gradient is estimated from a one-sided finite difference,

\begin{equation}
\label{eq:spgd}
    \nabla_{\mathbf{a}} J\left(\mathbf{a},\mathcal{E}\right)  \approx \left[ J(\mathbf{a} + \delta \mathbf{a}, \mathcal{E}) - J(\mathbf{a}, \mathcal{E})\right] \delta\mathbf{a}^{-1}
\end{equation}

The cost function is arbitrarily defined, allowing SGD methods to be applied based on the specific need. It can be an image quality metric such as Strehl ratio \cite{VorontsovSPGD97} for imaging systems, signal strength for free space optical communications \cite{weyrauchSPGCComms2004}, transmission coefficient for cold spot generation, or scattering matrix eigenvalue magnitudes for CPA state realization. Specific to the problem of controlling metasurfaces in microwave wavebands, energy efficiency in terms of bits-per joule is an attractive metric for wireless networks \cite{rodopluEnergyEfficiency2007}. Energy efficiency optimization using a reconfigurable metasurface has been proposed and simulated using both SGD and sequential programming in an open scattering environment \cite{huangRISWireless2019}. 

SGD methods work well in principle for controlling a metasurface. However, they begin to fail with coarse quantization, which limits the ability to tune both the size of the applied perturbation and the size of the step taken along the gradient. In the extreme case of a binary (1-bit) metasurface, applying a perturbation boils down to simply toggling or not toggling each element, so that for the $n^{\text{th}}$ element, $\delta a_n = \{0,1\}$. This leads to singularities in estimating the gradient (Eq. \ref{eq:spgd} ), as well as approximation errors with driving the solution along the gradient (Eq. \ref{eq:sgd}), since the resulting command must also be quantized to either 0 or 1. While metasurfaces can be manufactured with more bits of resolution for phase control, this increases complexity, cost, and power consumption considerably, making them less attractive for wide scale use. The capability of binary metasurfaces has been demonstrated many times; these devices can be expected to be utilized whenever power and cost are drivers for implementation. 

Since gradient based approaches are problematic with coarse quantization, dithering methods have dominated for wavefront control applications with binary tunable metasurfaces. We can modify the dithering technique in Eq. \ref{eq:dither} to use shaped or intelligent perturbations. When the algorithm is initialized, we do not know where the optimal commands are located with respect to the solution space. We would like to apply ``larger'' effective changes that induce highly diverse responses with large scale global changes. As the algorithm proceeds, effectively moving along the gradient, we want the changes to become ``smaller'', and more localized. In this manner, we are able to continue the optimization process without wasting trials on global changes that are less likely to improve the specific metric of interest. Finally, once the algorithm has converged, we would like to be able to make sure we are not stuck in a local minimum.

This shaped perturbation approach was demonstrated to successfully enable generating cold spots and realizing CPA states for a binary metasurface with 240 elements \cite{frazierWavefrontShaping2020}. In this case, the perturbations were ``shaped'' by changing the number of elements that were toggled (perturbed) each time the algorithm converged. The algorithm cycled through perturbations that toggled 120, 48, 24, 12, and then 6 elements, with a convergence criteria of $T = 30$ trials. This can also be thought of as a simple policy-iteration method of reinforcement learning \cite{bertsekasReinforcementLearningOptimal2019}. To ensure the solution was not stuck in a local minimum, the algorithm then entered a ``single element'' phase where three trial command vectors were generated at each iteration that toggled the individual element, the nearest neighbors of that element, and the diagonal neighbors of that element. Cold spots were generated with this technique and provided 4-40 dB of suppression over a 1 GHz frequency range and CPA states were realized and verified with power absorption ratios $\sim 10^6$ \cite{frazierWavefrontShaping2020}.

While dithering allows us to provide wavefront control in some capacity, a true wavefront reconstruction method is desired. Deep learning may provide a viable approach, as it has already been successfully demonstrated for general ill-posed inverse problems \cite{DengNeuralAdjoint2021}.

\section{Complex Scattering Environments}
\label{app:complexScattering}
 Large enclosed spaces, such as offices or compartments on ships or aircraft can act as ``chaotic'' reverberating chambers for short-wavelength electromagnetic waves \cite{drikasApplicationRandomCoupling2014}. These complex scattering environments contain universal fluctuations with statistics governed by random matrix theory (RMT) \cite{haakeQuantumSignaturesChaos2010}, as well as deterministic behavior from the system specific configuration of the ports and short orbits (i.e., prompt, direct paths) between the ports \cite{hemmadyUniversalImpedanceFluctuations2005,hartEffectShortRay2009,yehUniversalNonuniversalProperties2010}. We often characterize complex scattering environments by their scattering matrix, or $S$-matrix, which is a frequency dependent transfer function matrix containing the complex-valued reflection and transmission coefficients between the ports. While useful for describing the overall behavior, separating the universal and deterministic features is often difficult when working with the $S$-matrix.

An analytical approach known as the random coupling model (RCM), has been shown to accurately predict fluctuation statistics and allows separation of the universal and deterministic contributions in a simple additive manner \cite{zhengStatisticsImpedanceScattering2006a,zhengStatisticsImpedanceScattering2006}. Like RMT, the RCM leverages the random plane wave hypothesis, which asserts that the chaotic wave field is statistically equivalent to a random superposition of plane waves \cite{berryRegularIrregularSemiclassical}. The RCM is characterized by a single parameter, $\alpha$, that describes the losses in the system, and is supported by wealth of experimental validation data with chaotic microwave cavities \cite{hemmadyUniversalStatisticsScattering2005,hemmadyExperimentalTestUniversal2006,hemmadyUniversalPropertiesTwoport2006, yehExperimentalExaminationEffect2010,addissieExtractionCouplingImpedance2019}. The behavior in large, thee-dimensional enclosures has also been studied to understand these statistics and the potential impact of high power microwave (HPM) attacks \cite{drikasApplicationRandomCoupling2014,gilgilPredictionInducedVoltages2016}. 

The RCM works in the impedance domain to separate the universal contributions; conversion between impedance and scattering is handled through standard bilinear transformations \cite{pozarMicrowave2011}. In the RCM, the fluctuating impedance, $\bar{\mathbf{Z}}$, is defined as

\begin{equation}
    \bar{\mathbf{Z}} = j\text{Im}\{\mathbf{Z}_r\} + \text{Re}\{\mathbf{Z}_r\}^{1/2}\mathbf{\xi} \text{Re}\{\mathbf{Z}_r\}^{1/2}
\end{equation}

Here, $\mathbf{Z}_r$ is the radiation or free-space impedance of the ports and $\xi$ is the fluctuating or universal component, which is described by RMT \cite{zhengStatisticsImpedanceScattering2006a}. For lossless systems, $\xi$ is a Lorentzian distributed random variable. With loss, the distribution becomes much more complicated, but it is well suited to Monte Carlo simulations \cite{hemmadyStatisticalPredictionMeasurement2012}.

The RCM has been shown to apply to fading statistics in open systems, such as wireless communication paths, as well as closed systems, such as microwave cavities. \cite{yehFadingStatisticsCommunications2011,yehFirstprinciplesModelTimedependent2012}. A chaotic system is characterized by extreme sensitivity to initial conditions and is qualitatively different than an open one. In open systems, fading statistics are often modeled with empirically fit distributions \cite{simonCommunication2005}: the Rayleigh distribution when no line-of-sight path is present, the Rician distribution when a strong line-of-sight path is present, or the K distribution for propagation over the ocean \cite{wardKDistribution2013}. The limiting cases of Rayleigh and Rician distributions are captured by the RCM with the $\sigma$ parameter related to the loss parameter, as $\alpha = \left(8\pi\sigma^2\right)^{-1}$, and the $\nu$ parameter equal to the magnitude of the short orbits \cite{yehFadingStatisticsCommunications2011,yehFirstprinciplesModelTimedependent2012}.

Operating in a rich scattering environment presents an additional set of challenges in comparison to an open environment. In addition to the fundamental difference in the character of the fluctuations, in the semi-classical case or short wavelength limit, we can look at the behavior of ray trajectories. Specifically, we are interested in the change in ray trajectories in response to a change in the metasurface configuration. In an open system, there is a single ray (or ray bundle) that is observed by the sensor, with at most a single reflection off the metasurface. In a chaotic system, that single ray will reflect off multiple walls and obstacles in the cavity and possibly off the metasurface itself multiple times before reaching the sensor. This creates a cascading effect, so that the wavefront at the sensor is a combination of constructive and destructive interference of the multiple rays. The effect of these multiple interference paths is highly dependent on the configuration of the cavity. For an open system, a wavefront reconstruction approach is only dependent on the geometry between the sensor and the correcting device and is invariant to environmental changes (provided the disturbances remain within the dynamic range of the sensor and corrector). Wavefront reconstruction is therefore very robust for an open system. For a chaotic system however, small environmental changes can cause a wavefront reconstruction technique that was previously successful to no longer be viable, so that being ``close" is not good enough.

To build environmental models for simulations, we often need to make assumptions or approximations for simplicity or computational tractability. We also need to ensure that these assumptions are valid for the environments that are being modeled. Otherwise, the models may neglect potentially significant effects. We will outline some of the most problematic simplifying assumptions here.

The first simplifying assumption often made is that the channels are assumed to be perfectly known by the transmitter (base station), so the only uncertainty in the environment is random thermal noise at the receiver(s). In complex scattering environments, there is always uncertainty, which can be significant. In addition, inside chaotic cavities, measured scattering responses with the same initial conditions will diverge over time, a phenomenon known as scattering fidelity decay \cite{SchaferExperimentalFidelity2005,taddeseSensorBasedExtending2009,taddeseSENSINGSMALLCHANGES}, which means that perfect knowledge of a complex scattering environment has a finite lifetime. This lifetime can be several days in controlled conditions, but is sensitive to temperature and humidity and will be reduced in scenarios such as dense urban environments. Having only partial knowledge of the system limits deterministic control approaches and encourages learning algorithms. Scattering fidelity has an impact here as well; as the scattering responses start to diverge, any machine learning algorithm will require periodic retraining.

The second simplifying assumption often made is that the equivalent channel matrix, is assumed to be invertible, so the inverse problem is well-posed. Complex scattering environments generally contain short orbits, or prompt direct paths, that are persistent across measurements \cite{hartEffectShortRay2009,yehExperimentalExaminationEffect2010}. These short orbits induce correlations that are difficult for simple machine learning approaches to unwrap and typically lead to ill-posed inverse problems. Excluding multi-path reflections and short orbits can overestimate the performance of a given algorithm.

The third simplifying assumption often made is that all the propagation paths are assumed to have a single reflection off the metasurface, so direct line-of-sight and multi-path trajectories are not included. Channel fading is then modeled with Rayleigh amplitude statistics. As stated previously, in real-world systems, strong direct line-of-sight paths induce Rician statistics and the presence of multi-path reflections drives statistics that are governed by random matrix theory (RMT) \cite{yehFadingStatisticsCommunications2011,yehFirstprinciplesModelTimedependent2012}. Neglecting these statistics can lead to algorithms that are not properly tuned. The longer tails in the distributions lead to large amplitude signal spikes that can degrade imaging performance or disrupt signal processing algorithms.

Testing and verification is often done in anechoic chambers to remove the environment and capture only the impact of the reconfigurable metasurface. Anechoic chambers are very good at covering up emissions problems, which become immediately apparent in reverberation chambers \cite{hollowayReverberationChamberTechniques2012}. A complex scattering system is reverberant in nature, so mismatches that seem negligible in anechoic chambers may be significant in real world environments.

Finally, the metasurface is often assumed to have infinite phase resolution, so quantization effects are not included. Most commercially available metasurfaces have a single bit of control, though custom devices with 2 bits of control are becoming available \cite{liTransmissionType2BitProgrammable2016,zhangSpaceTimeCoding2018, shaoDielectric2bitCoding2019,iqbalDualband2bitCoding2019}, which means quantization effects are important and likely significant. In addition, as discussed in Appendix \ref{app:wavefront_reconstruction_control}, coarse quantization can cause gradient based controllers to fail, so neglecting quantization effects may lead to poorly performing real world controllers. The metasurfaces are also idealized, with identical responses across all elements. In real devices, manufacturing defects produce nonuniformities between the phase at each element, and the metasurface may also include uncontrollable losses or gain.

Caution should be taken when applying any of these simplifying assumptions. Otherwise, they can overestimate the performance or install a false sense of confidence in a particular approach.

A final characteristic of complex scattering environments to describe is coherent perfect absorption (CPA). CPA is a special state of the scattering matrix where an eigenvalue is driven to 0 and the electromagnetic energy for the corresponding eigenvector is completely absorbed inside the scattering system \cite{chongCoherentPerfectAbsorbers2010,chenPerfectAbsorptionComplex2020}. While practical applications are still being developed, research has demonstrated that a high fraction of the power was absorbed by the target in a CPA demonstration using a tuned absorber embedded in a lossy environment \cite{chenPerfectAbsorptionComplex2020}. This work also showed that the target absorbed virtually nothing in the ``anti-CPA" state, demonstrating a high degree of control over absorption by a specific target in a CPA scenario. An interesting future application is to utilize a generalized Wigner-Smith operator \cite{ambichlGeneralizedWigner} to apply a high absorption fraction to a target with a modulated impedance or loss. 

Full coherent multi-channel CPA is more complicated than single channel perfect absorption. However, one advantage is the increase in delivered power by a factor of $N$, where $N$ is the number of channels. This is a significant gain and worth the difficulty of additional phase and amplitude control.

\section{Cavity and Experimental Configuration}
\label{app:cavity}
For the cavity discussed in the main paper, a line-of-sight block is used to obstruct the direct transmission path from port 2 to both ports 1 and 3. Also, ports 1 and 3 can be driven either independently or collectively with a relative phase shift provided by a NARDA phase shifter.

In the frequency range of interest, 3-4 GHz, the Weyl formula \cite{hillCavities2009} predicts approximately 8524 resonant modes of the complex enclosure and the measured quality factor of the cavity is roughly $5.5 \times 10^3$ \cite{frazierWavefrontShaping2020}. A resonance mode width is then $\sim$ 5 times greater than the mean mode spacing, which means there is some local overlap between modes.

\section{Metasurface Binning}
\label{app:metasurface_binning}
The complexity of the cavity scattering responses combined with the enormous number of possible metasurface command configurations ($2^{240}$) means the direct development of a deep learning network for the full space of 240 elements is overly ambitious. To simplify the problem, we reduced the number of degrees of freedom of the metasurface by binning together neighboring pixel elements, or grouping them together so that each element in a group is always commanded with the same value. Binning the metasurface elements reduces the total number of elements that must be determined and strengthens the relative change in cavity scattering parameters when driving a single effective element. Binning also promotes generality, as a metasurface with smaller elements can always approximate one with larger elements. This provides the first major novel aspect of our approach and allows us to explore the use of deep learning models in simpler configurations before working our way up to the more difficult cases. We used 4 different metasurface binning configurations, as shown in Fig. \ref{fig:metasurface_binning} and discussed in Section S4 of the supplemental material, to progressively decrease the number of elements. 

\begin{figure}
\includegraphics[width = 8.6 cm]{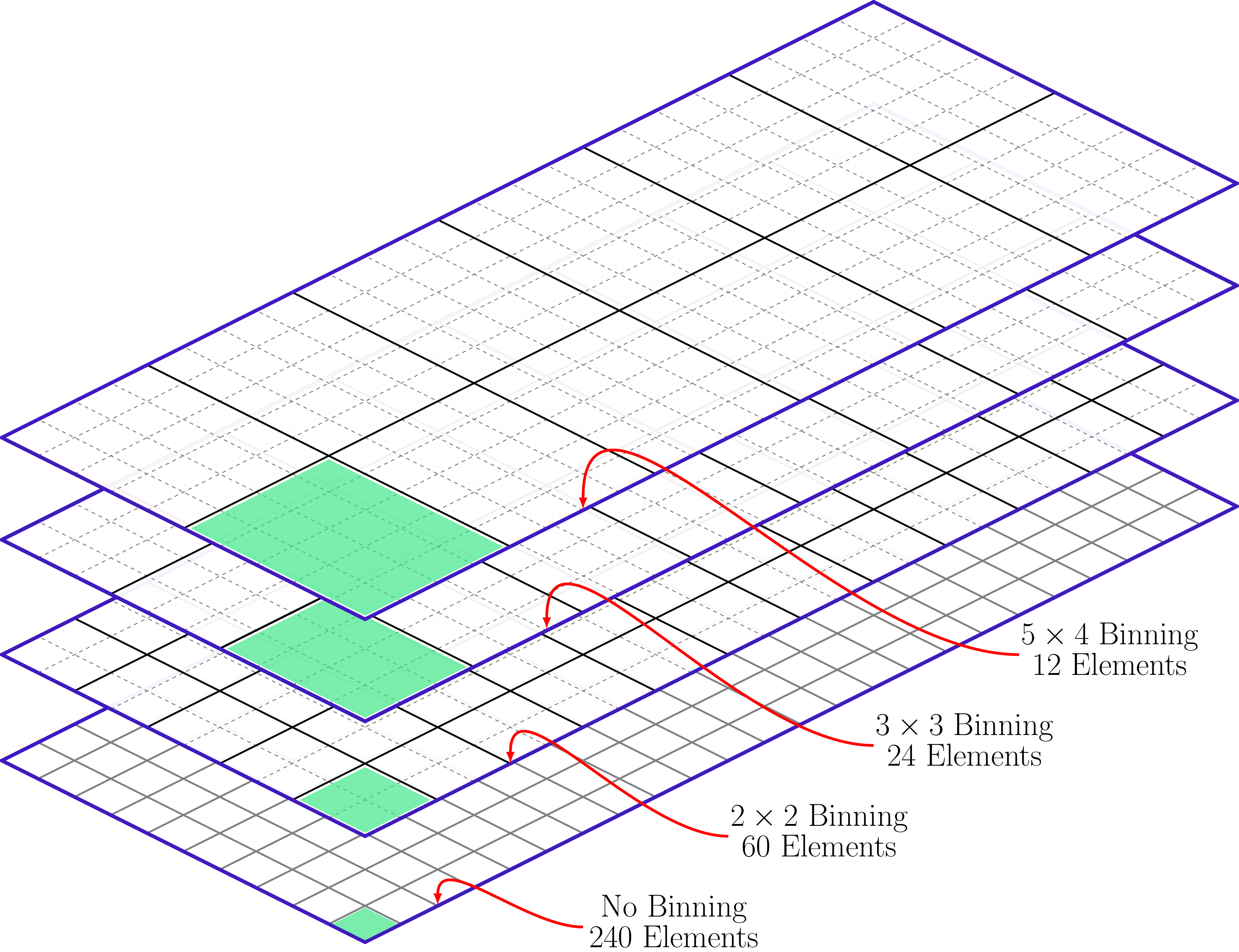}
\caption{\label{fig:metasurface_binning} \bf Metasurface binning configurations. \normalfont Binning configurations showing the relationship between the various options. The shaded green region identifies a single effective element for the specified configuration and the thin gray lines show the layout of the unbinned elements. With no binning, there are 240 elements, binning into groups of $2\times2$ yields 60 elements, binning into groups of $3\times3$ yields 24 elements, and binning into groups of $5\times4$ yields 12 elements. For the $3\times3$ binning case, the bottom row of elements consists of a $4\times3$ group so that all the elements are utilized.}
\end{figure}

\section{Data Preparation and Collection}
\label{app:data_prep}
A major concern with deep learning is the amount of data required for training, which grows with the complexity of the problem being solved. To work within the constraint of reasonable training time, we wish to limit the number of data sets that must be collected. Therefore, acquiring good training data is of critical importance to ensure we cover the full range of possible responses. As found in earlier work \cite{frazierWavefrontShaping2020}, a diverse set of measurements requires variations in the number of active elements, spatial frequencies of active elements, and local groupings of active elements. Therefore, we utilized a random biased coin toss approach with the bias itself a uniformly distributed random number to assign values to the elements for training data generation. To speed up operation as much as possible, the microwave network analyzer was configured to only provide $S_{21}$ measurements vs. frequency. With averaging disabled, collecting 4,000 sets of data took a little under an hour and a half, while collecting 10,000 sets of data took roughly 3.5 hours. Training was performed on a computer running Ubuntu 20.04 equipped with an NVIDIA RTX 3080 GPU, and took roughly 30 minutes for 10,000 training sets.

For the initial experiment, we collected 4000 sets of data in each of the specified binning configurations; the $5\times 4$ configuration allows 4096 unique metasurface combinations, so we collected 4096 sets (covering all possible combinations) in that case. With the exception of the $5\times 4$ binning configuration, the number of sets collected was far smaller than the number of possible configurations of the metasurface. As discussed in Section \ref{app:2x2}, we found we needed 10,000 sets of data for the $2\times2$ binning case.

\section{Deep Learning and Neural Network Layers}
\label{app:deep_learning}
The concept of depth in deep learning comes from complexity theory as defined for circuits, with  depth being the longest path from an input to an output \cite{bengioDeepLearningRepresentations2012}. The number of potential paths or ways to reuse features grows exponentially with depth, which leads to progressively more abstract features \cite{bengioDeepLearningRepresentations2013,montufarAdvancesinNeural2014}. Depth is therefore an important characteristic of a network to leverage as it enhances the expressive power of the network and allows it to learn a rich, hierarchical feature representation \cite{simonyanVeryDeepConvolutional2015,raghuExpressivePower2017}. In physics, deep learning has repeatedly been shown to be unreasonably effective for extremely complicated problems \cite{sejnowskiUnreasonableEffectivenessDeep2020}.

A deep network is divided into layers, with the interior layers often referred to as hidden, as they are unobservable from the input or the output. Networks come in many different shapes and sizes; no single type is optimal for all problems, a consequence of the ``No Free Lunch'' theorem \cite{wolpertNoFreeLunch1997}. The networks described in this paper utilize 4 different types of layers: 1) Dense, linear, or fully connected layers characterized by the number of neurons. The output is a linear combination of the inputs; 2) Convolutional layers characterized by the number of filters and the length of the kernel (see Section Appendix \ref{app:1dConvolution}). The output is the result of convolving the inputs with the kernels; 3) Pooling layers characterized by the pool size. The output is either the maximum or average value over a sliding window of width given by the pool size. These layers serve to reduce the size of the feature map and help ensure the learning process is position invariant; and 4) Dropout layers characterized by the drop out rate. Dropout layers randomly set the specified percentage of inputs to $0$ at each iteration in the training process, providing coarse regularization and simplifying the model.

Identifying the optimal deep learning network topology for a given binning configuration took a significant amount of time to iterate over many potential designs, e.g., for our computing resources, often several weeks, and was performed off-line. Once the deep learning network architecture was determined, we switched to an on-line, closed loop configuration where the data collection and training processes were separated by a few hours rather than days or weeks. The determined metasurface commands were directly applied to the metasurface and the resulting $S_{21}$ responses were measured by the network analyzer, closing the loop. The on-line configuration also serves as a ``field-test'' for the deep learning network, further validating it against data not seen during training, as well as testing performance against potential small variations in measurement noise and the scattering configuration of the cavity itself. 

\section{1D Convolution}
\label{app:1dConvolution}
\begin{figure}
\includegraphics[width = 8.6cm]{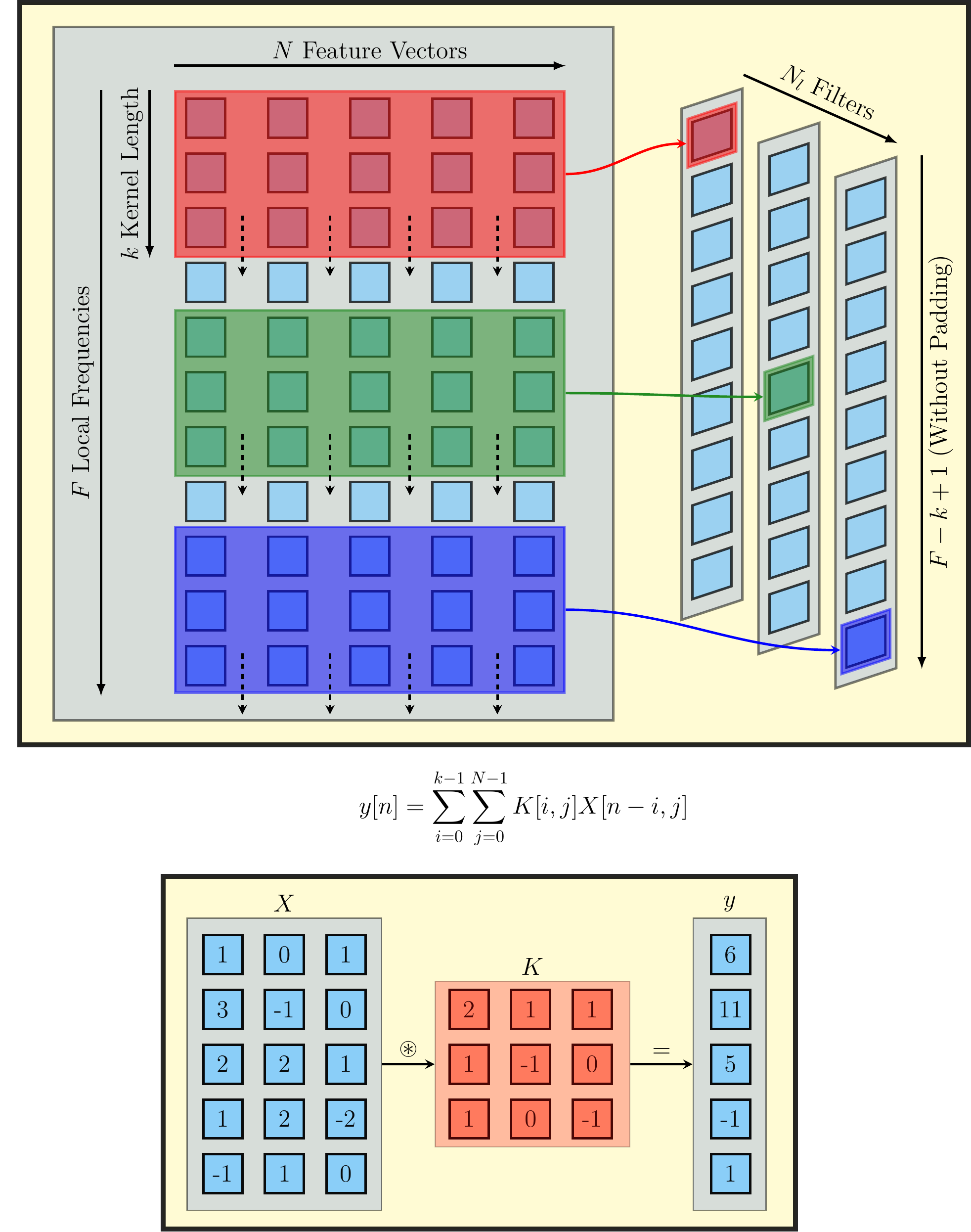}
\caption{\label{fig:1dcnn} \bf 1D Convolution with Multiple Features. Top: \normalfont Graphical representation showing the $N$ feature vectors and $F$ local frequencies processed by 3 different filters with kernel length $k$. The kernel only moves along a single dimension (vertically) even though the data is represented in a 2D format. Each position of the kernel results in a single point in the output vector which has length $F-k+1$ if zero padding is not used and length $F$ if padded as described in the text. The weights for each of the $kN_l$ elements of the kernel are computed collectively, but can be different. \bf Bottom: \normalfont Numerical example with 3 features containing 5 points each convolved with a kernel of length 3. The input data is zero padded with a row of zeros at the top and bottom and the outputs for the 5 central rows are kept.}
\end{figure}

An aspect that is not well understood outside of the signal processing community is how convolutional layers are implemented for inputs containing multiple features. In signal processing, the feature dimensionality is referred to as the number of channels and is sometimes defined as the width or the depth of the data. This arises from color image processing with 3 color channels for red, green, and blue. To perform the convolution over the desired dimension and ensure all the features are captured, the convolution kernel is multidimensional as shown in Fig. \ref{fig:1dcnn}. For a specified kernel length, $k$, the size of the kernel for a 1D convolutional layer with an input containing $N$ features is $k\times N$. The kernel will only be shifted along a single dimension, the local frequency window in our case, but will contain optimized weights for each element. This means that the number of trainable parameters for a 1D convolutional layer scales as $kN$, not just $k$. For an input data set $X$, the output, $y$, of the convolutional layer with kernel $K$ is given by

\begin{equation}
    y[n] =\sum_{i=0}^{k-1}\sum_{j=0}^{N-1} K[i,j]X[n-i,j]
\end{equation}

For our purposes, we will zero pad the input data by appending $(k-1)/2$ rows of zeros to either end and keeping the central part of the result, so the number of points along the convolution dimension is constant in the output. By designing a convolutional layer consisting of $N_l$ filters, there will be $N_l$ such outputs or new features for the next layer.

\section{Network Training Setup}
\label{app:network_training}
For training the networks, the data was split into 75\% training data and 25\% validation data. The validation data is used to score the performance after each training run and is not used during the training process itself, so that validation is unbiased. The data was randomly shuffled prior to splitting and each network was trained several times to ensure results were in family and that the training process was unbiased as well. 

To score the performance, we need to define metrics for loss and accuracy. The loss function was selected as mean absolute error to emphasize outliers in the data and we define accuracy as the fraction of sets of commands that were predicted without error. To clarify the difference, the loss function is defined as the average of the sum of the absolute value of the true commands, $T_j$, subtracted from the predicted commands, $P_j$, for set $j$, computed over $N$ sets of $M$ elements.

\begin{equation}
L = \frac{1}{NM}\sum_j\left|P_j - T_j \right|
\end{equation}

The loss function is then computed on a per element basis and tells us how close the prediction was on average for each element. The output of the network is floating point rather than binary, so the loss function does not necessarily provide an indication of the total number of incorrect predictions. The accuracy metric is defined as the percentage of sets that were predicted without a single error. It is evaluated on a per set basis and explicitly uses the rounded output (0 or 1) from the network. Because accuracy is computed on a per set basis, it is dependent on the number of elements in a command set and provides a more conservative estimate of performance for the various binning configurations. Accuracy is also more volatile, especially when the loss function is large. The loss function is continuous and more appropriate for training where we need to compute a gradient, while accuracy is a better metric for scoring the overall performance.

The networks were trained for 100-200 epochs using stochastic gradient descent (SGD) with momentum. The basic SGD algorithm has potential problems with pathological curvature, or narrow ravines, which are common around local optima, and the response tends to oscillate back and forth across the ravine. To address this, we can use momentum \cite{sutskeverImportance2013}, effectively forgetting a portion of the previous gradient. Momentum can be thought of as a very coarse approximation of the curvature or 2nd derivative. To accelerate the training, we explicitly use Nesterov momentum \cite{nesterovMethod1983}.

The networks were trained in batches, meaning multiple data sets were evaluated at each iteration prior to updating the weights. This allows several samples of data to be processed simultaneously so that the effect of changes in the weights are observed over multiple sets of data, improving robustness and desensitizing the response to noise \cite{smithDisciplinedApproachNeural2018}. A batch size of 100 was used by default. 

To prevent the networks from simply training on noise, we introduce an additional regularization step on the loss function. By enforcing an $\mathcal{L}_2$ regularization scheme, the regularized loss function $L^*$ is computed from the loss function $L$, and the vector of weights for the current iteration, $\mathbf{w}_i$, as $L^* = L + \lambda ||\mathbf{w}_i||^2$. The value $\lambda$ is referred to as a weight decay. The learning rate, $\gamma$, is the step size along the gradient, so the weights are incremented at each iteration as

\begin{equation}
\mathbf{w}_{i+1} = \mathbf{w}_i - \gamma \nabla_{\mathbf{w}_i}L - 2\gamma \lambda \mathbf{w}_i
\end{equation}

Finally, the learning rate is stepped down when the loss function plateaus, which allows the network to continue learning when it stalls due to the rate being too high.

\section{Complex Network Layers and Existing Deep Learning Frameworks}
\label{app:complex_layers}
With complex values, the mechanics of a network layer are the same as for the real-valued counterpart but they incur four times the computational cost due to having both real and imaginary components as well as the cross-terms. Our initial deep learning implementation leveraged Keras \cite{cholletKeras2015} and TensorFlow \cite{abadiTensorFlowLargeScaleMachine2016}. These provide an excellent, high level framework that is very easy to use. Unfortunately this ease of use complicates things when attempting to develop custom complex-valued modules. Complex dense layers are straightforward to implement, but batch normalization, convolution, and recurrent layers are not. While there are repositories with complex deep networks containing some of these modules in Keras \cite{trabelsiCode2018} and Caffe \cite{virtueCode2018}, they are not actively maintained and are not formally supported by the frameworks. In the case of Keras, changes to the way the backend is handled in the most recent version (v2.4) mean that the complex library \cite{trabelsiCode2018} is no longer functional and would require significant modification to bring up to date.

This leads us to utilize PyTorch \cite{NEURIPS2019_9015}, another deep learning framework. The interface to PyTorch is lower level than Keras, which means it requires more knowledge of Python to use effectively, but that it is also easier to implement custom modules. In addition, there is an open source complex library written by Sebastien Popoff \cite{popoffComplexPyTorch2019} that includes complex versions of dense, convolutional, and batch normalization layers. We were able to utilize this library with only minor modifications to the batch normalization implementation to handle our multiple feature data sets.

\begin{figure*}
\includegraphics[width = 17.2cm]{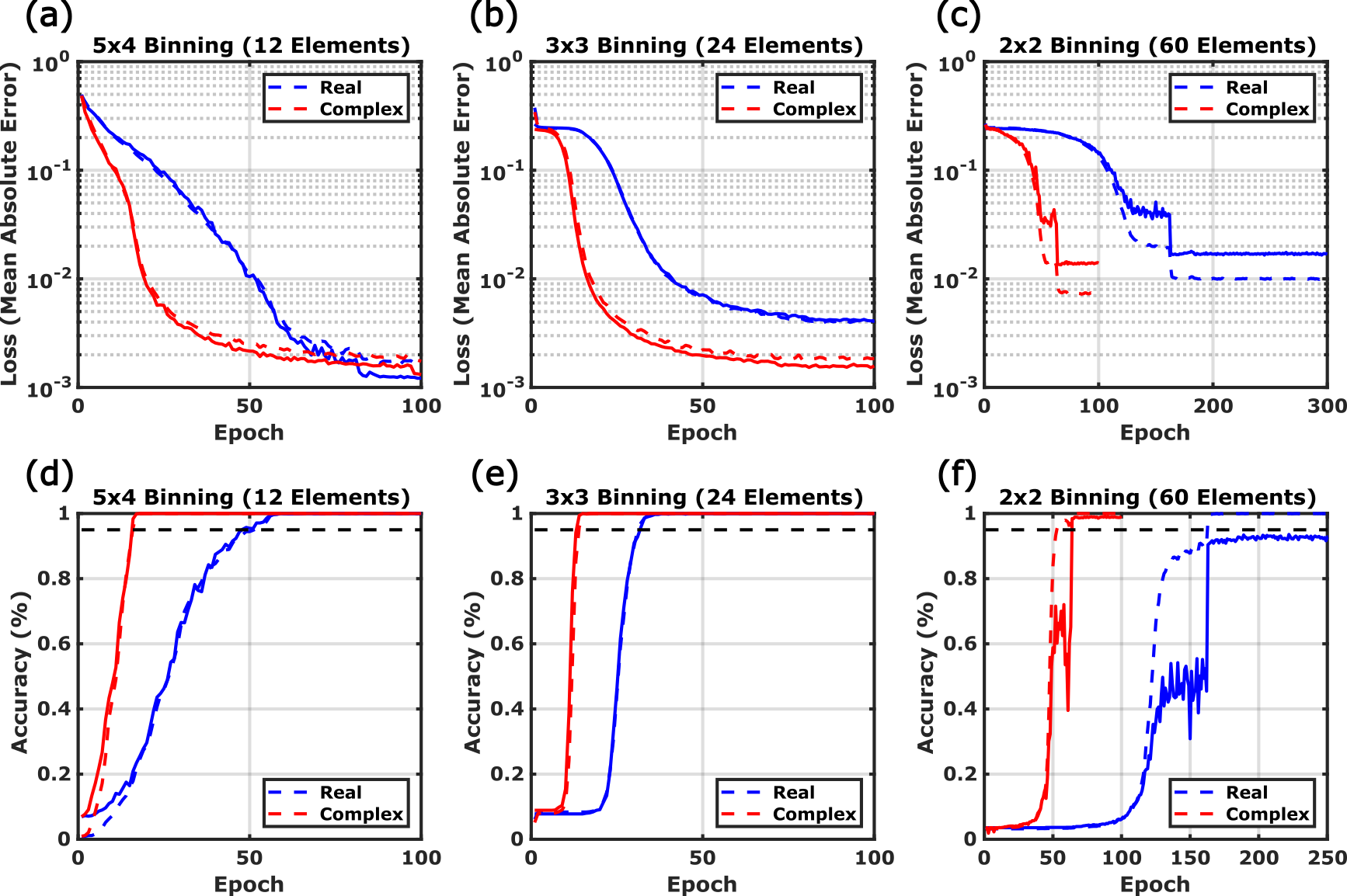}
\caption{\label{fig:complexImpact} \bf Impact of using complex network layers. a) \normalfont through \bf c) \normalfont Evolution of the loss function (mean absolute error) for the $5\times4$, $3\times3$, and $2\times2$ binning cases. \bf e) \normalfont through \bf f) \normalfont Evolution of the accuracy for the $5\times4$, $3\times3$, and $2\times2$ binning cases. The blue lines indicate real-valued network layers, the red lines indicate complex-valued network layers, the dashed lines indicate the training set, and the solid lines indicate the validation set. The dashed black line on the accuracy plots indicates 95\% accuracy. Training was performed for 100 epochs in all cases except for the $3\times3$ binning case with real-valued network layers, which was trained for 300 epochs. In each case, training with the complex-valued layers converged faster than training with the real-valued layers. For the $2\times2$ binning case, the complex-valued network achieved higher accuracy for the training set (99.2\%) than the real-valued network did (93.6\%).}
\end{figure*}

Figure \ref{fig:complexImpact} shows the impact of using complex network layers. In each of the panels, the blue plots indicate results for a real-valued network while the red plots indicate results for a complex-valued network. In addition, the solid lines show results for the validation set while the dashed lines show the results for the training set. The complex-valued networks all converged faster than the real-valued networks and the complex-valued network achieved better accuracy on the training set than the real-valued network did for the $2\times2$ binning case. The only differences in training were for the real-valued network in the $2\times2$ binning case. The two differences here were that 1) the "patience" parameter, or number of epochs to wait before reducing the learning rate had to be increased significantly because the training converged slowly even with an aggressive learning rate; and 2) the number of epochs had to be increased from 100 to 300 in order to capture the converged model. The increased patience parameter leads to a longer period of oscillations in the validation set loss function.

The acceleration in training comes with a caveat in that the overall computational time for the purely real-valued deep learning network is still less than that of the complex-valued deep learning network. Complex-valued layers increase the computation requirements for multiplication and convolution by a factor of 4 to handle the real and imaginary terms as well as the cross-terms. In addition, highly optimized and efficient implementations of purely real-valued layers are readily available through the NVIDA CUDA deep neural network library (cuDNN), but are not available for their complex-valued counterparts.

\section{Network Architecture for Sequential Layers}
\label{app:network_architecture}
\begin{figure*}
\includegraphics[width = 17.2cm]{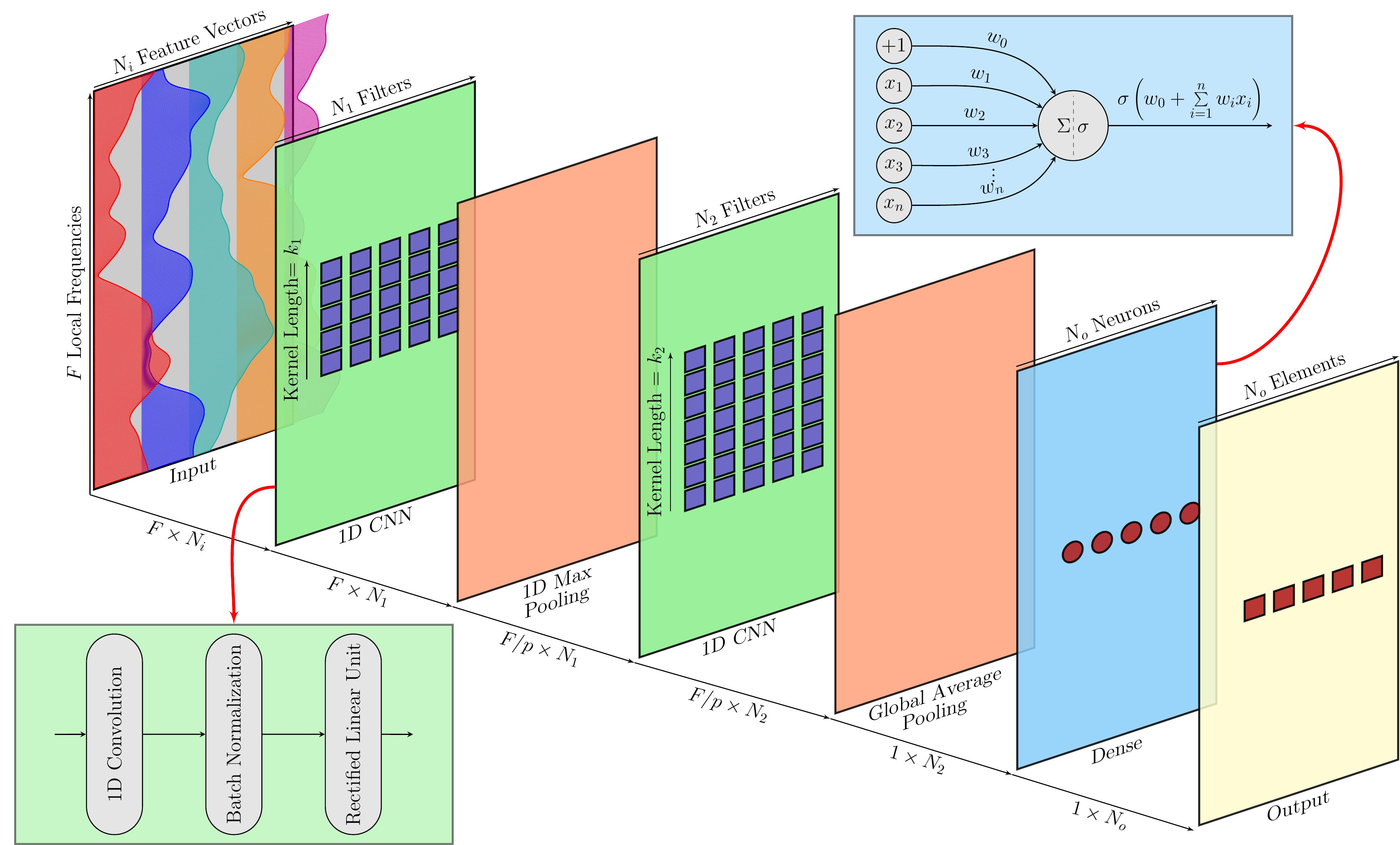}
\caption{\label{fig:sequential_layer_structure} \bf Sequential network layer architecture. \normalfont The input layer consists of $N_i$ feature vectors containing $S_{21}$ measurements in a local 10 MHz window of $F$ points, for a total size of $F\times N_i$. This is followed by a series of 1D convolutional layers defined by the number of filters and the kernel length. Each convolutional layer includes a 1D convolution, a batch normalization to keep the distribution statistics constant throughout the network, and a rectified linear unit activation function. Interspersed with the convolutions are 1D max pooling layers defined by the pool size, $p$, that serve to reduce the dimensionality of the local frequency window. The output stage consists of a global average pooling layer to further reduce dimensionality and convert the complex-valued signals to magnitude, followed by a dense or fully-connected layer to ensure the correct number of outputs. The dense layer produces outputs that are linear combinations of the outputs from the global average pooling layer, and is followed by a sigmoid activation function to approximate binary values at the output.}
\end{figure*}

Figure \ref{fig:sequential_layer_structure} presents the generalized architecture used for sequential layers. The input consists of $N_i$ feature vectors containing the local 10 MHz windows with $F$ points in each vector. 1D convolutional neural network (CNN) layers with $N_l$ filters and a kernel length of $k_l$ at the $l^{\text{th}}$ layer perform the feature extraction. As shown in the lower inset, each CNN layer includes a 1D convolution followed by a batch normalization and a rectified linear unit (ReLU) activation function. The batch normalization is used to ensure the distribution of the data (mean and variance) remains relatively constant throughout the network; changing distributions between layers induces internal covariate shift and leads to convergence issues during training \cite{ioffeBatchNormalizationAccelerating}. The ReLU activation function is used in virtually all deep learning networks as it does not experience a vanishing gradient due to saturation, and leads to expedited convergence and generally better solutions than sigmoid like functions \cite{heDelvingDeepRectifiers2015}.

The CNN layers are grouped together to form stages and are interspersed with 1D max pooling layers defined by the pool size, $p$. Also included are dropout layers for regularization, which are not explicitly shown in Fig. \ref{fig:sequential_layer_structure}. The output stage contains a global average pooling layer that averages along the $F$ dimension, reducing the feature maps to a single dimension. It also converts the complex-valued signals to magnitude and is followed by a dense layer that provides the correct number of outputs, $N_o$. As shown in the upper inset, the output of the dense layer is a linear combination of the outputs from the global pooling layer, with a sigmoid activation function used to clip the output between 0 and 1.

\section{\label{app:bin5x4}Offline Training Results for 5 x 4 Binning}
When using $5\times 4$ binning, the metasurface is effectively partitioned into 12 elements, for 4096 possible sets of commands. We measured all 4096 combinations, and the 75\%/25\% split yielded 3072 sets for training and 1024 sets for validation. The batch size was set to 64 as a result. A purely sequential deep network was utilized, following the layout given in Fig. \ref{fig:sequential_layer_structure}. Four CNN layers, a max pooling layer, and a dropout layer were combined into a stage. Four stages were then used, with the output provided by a global average pooling layer and a dense layer with a sigmoid activation function. 

Initial experiments used purely real-valued deep learning layers, in which case the global average pooling layer only provided dimensionality reduction. We were able to establish excellent prediction performance, regularly achieving $<$ 10 total prediction errors over the validation set or $>$ 99\% accuracy after training for 500 epochs. When switching to complex-valued layers with the same architecture, we were able to regularly achieve perfect prediction (100\%) over both the training and validation sets after training for fewer than 100 epochs. This improvement demonstrates that the complex-valued deep learning network is able to exploit phase as well as amplitude to better fit the relationship between metasurface commands and transmitted power.

\begin{figure*}
\includegraphics[width = 17.2cm]{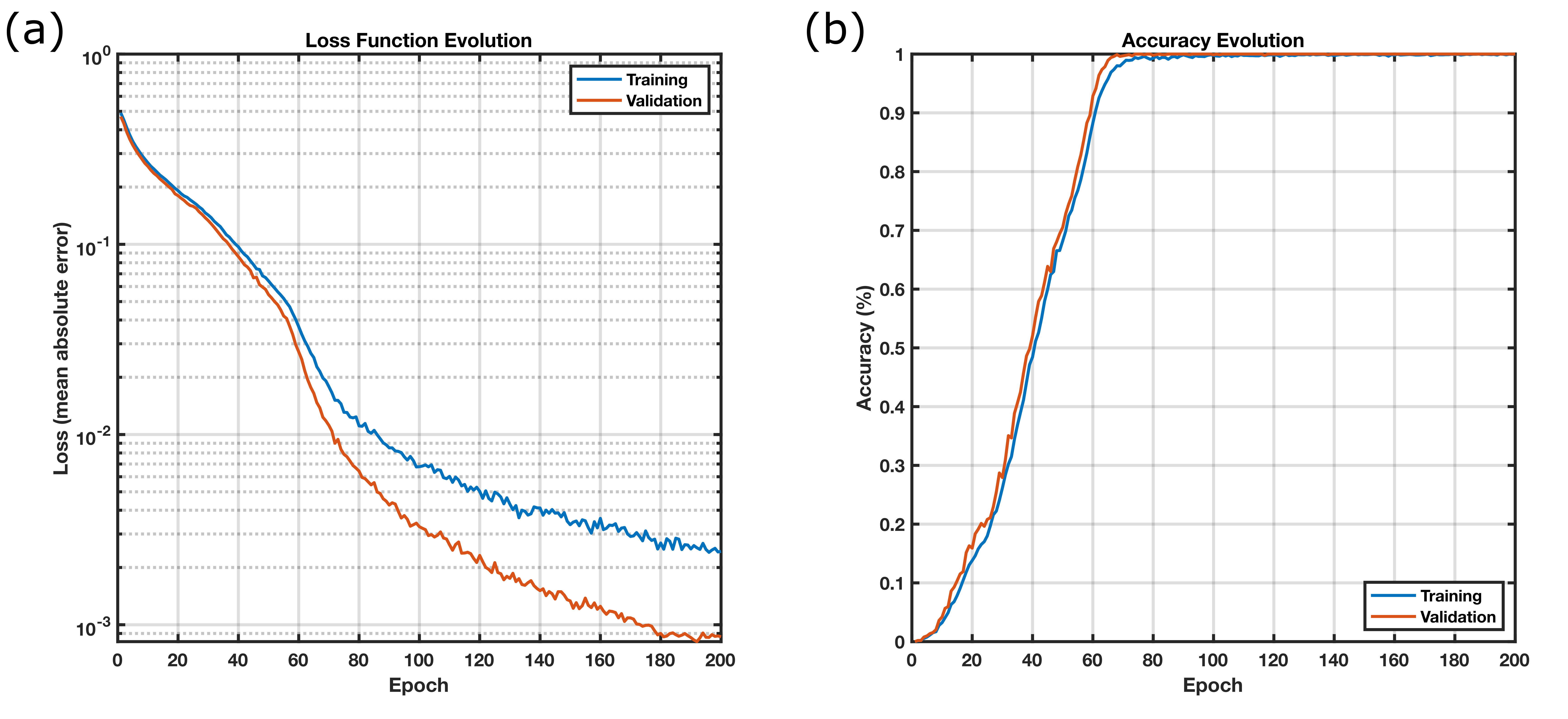}
\caption{\label{fig:performance_5x4_binning} \bf Deep learning performance with complex-valued layers for 5x4 Binning. \normalfont \bf (a) \normalfont Evolution of the loss function for the training and validation sets over 200 epochs. The loss function measures the average prediction error per element and provides an estimate of the confidence in the prediction. \bf (b) \normalfont Evolution of the accuracy for the training and validation sets over 200 epochs. Accuracy provides the relative number of sets of commands that were predicted without error, and shows that perfect prediction was achieved on both the training and validation sets in less than 100 epochs. The loss function continues decreasing after the accuracy saturates at 100\% because it is continuous and evaluated on the floating point predicted values and the decrease indicates the network is still learning and improving its estimate.}
\end{figure*}

The training results for the $5\times4$ binning case with complex-valued layers are shown in Fig. \ref{fig:performance_5x4_binning}. Figure \ref{fig:performance_5x4_binning} (a) shows the loss function evolution for the training and validation sets while Fig. \ref{fig:performance_5x4_binning} (b) shows the accuracy evolution. The performance is excellent, achieving perfect prediction over both the training and validation data sets. Note that the loss function continues to improve even after the accuracy saturates at 100\%. This is because the loss function is continuous and computed using the floating point predicted values rather than the rounded binary values; it shows that the network is still learning and continuing to increase its confidence in the prediction.

\section{Receptive Field and Inception Module}
\label{app:receptive_field}
 The receptive field of a CNN defines the number of points in input space that contribute to the result at a single point in a given layer. Our CNNs use zero padding to keep the output size fixed, and the stride and dilation are always set to 1. This means the receptive field at any layer, $r_l$, is given by a simple recursive equation dependent on the receptive field at the previous layer, $r_{l-1}$, and the kernel length of the current layer, $k_l$ \cite{araujo2019computing}.

\begin{equation}
r_l = r_{l-1} + k_l - 1
\end{equation}

\begin{figure}
\includegraphics[width = 8.6cm]{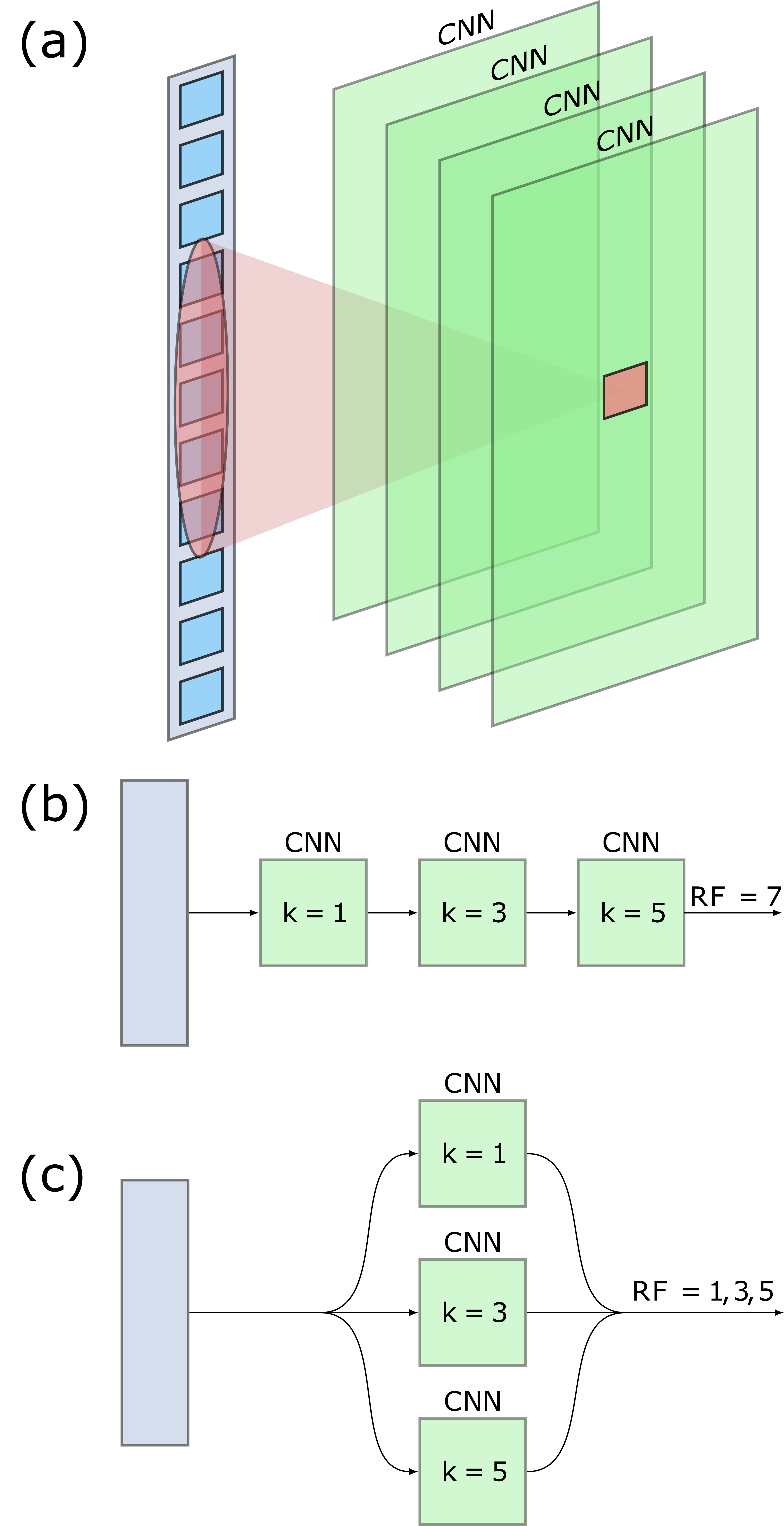}
\caption{\label{fig:receptiveField} \bf Receptive Field. \normalfont \bf (a) \normalfont The receptive field of a convolutional neural network (CNN) layer indicates the number of points in input space that contribute to a single point at a given layer. \bf (b) \normalfont For a purely sequential architecture, the receptive field increases monotonically. \bf (c) \normalfont A parallel architecture with concatenation produces multiple receptive fields with each available for subsequent layers, promoting sparsity in the representations. }
\end{figure}

As shown in Fig. \ref{fig:receptiveField}, the receptive field for a sequential architecture grows monotonically with depth, with each layer only seeing the receptive field from the preceding layer. An architecture that utilizes parallel branches along with concatenation conserves the intermediate receptive fields, making them available for all subsequent layers and introduces width as well as depth to the network and providing the motivation for the inception module.

An inception module is designed to promote sparse feature representation using available dense components \cite{szegedyGoingDeeperConvolutions2015} and works by optimizing the receptive field coverage of a convolutional network. The receptive field is the number of points in input space that contribute to a point at a given layer of the deep learning network and is described in detail in Section S11 of the supplemental material. Through the use of parallelization and concatenation, the receptive field sizes at a layer are conserved for subsequent layers to utilize, extracting features through the width of the deep learning network as well as its depth.

The original inception module was developed for image processing and operates in a true 2D space, with full 2D convolutions. It uses 4  parallel paths with CNN layers containing unit length kernels for buffering and conditioning, along with 3 and 5 sample length kernels for feature extraction, and a max pool layer to improve performance \cite{szegedyGoingDeeperConvolutions2015}. There have been several variations of the inception module; however, none operate in the pseudo-2D space we desire. For our ``images'', the frequency spacing along columns is the resolution of the network analyzer (31.25 kHz), while the frequency spacing along rows is the separation between local windows (10 MHz). The difference in sampling means we need to treat the rows and columns accordingly and avoid traditional 2D convolutions that assume uniform sampling. Therefore, we modified the general architecture of the inception module to perform 1D convolutions over the 10 MHz local frequency windows. The 1D convolutional filters then extract local features over the 10 MHz windows, while the relationship between the filters acts as a dense or fully connected layer, extracting global features over the full 1 GHz measurement window.

\begin{figure}
\includegraphics[width = 8.6cm]{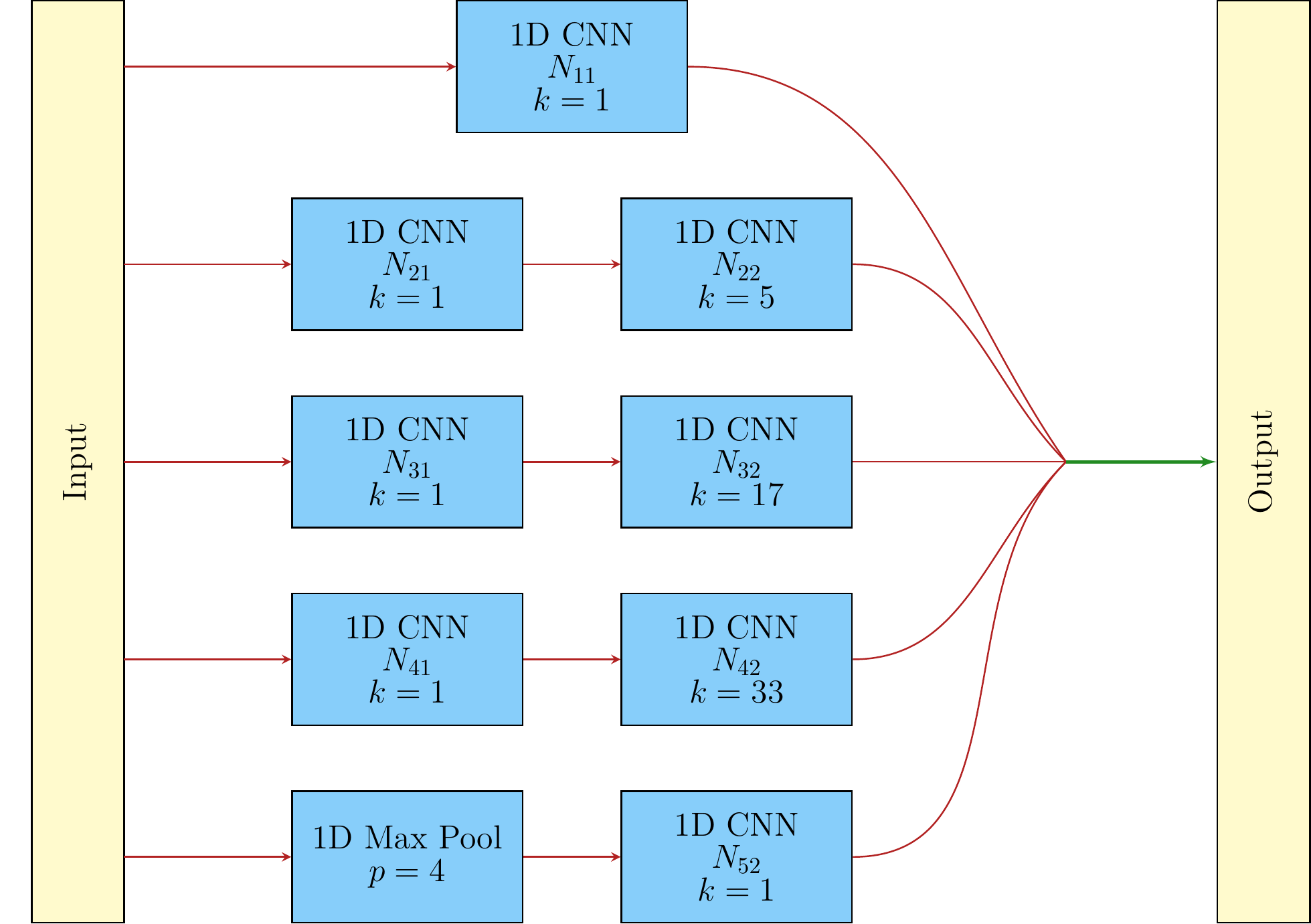}
\caption{\label{fig:modified_inception_module} \bf Terrapin Module Architecture. \normalfont Five parallel branches with 8 1D convolutional neural network (CNN) layers and a max pool layer are used in the module. The module operates on the pseudo-2D data format discussed in the text, and the input layer can ingest either the raw measured $S_{21}$ parameters or the outputs from a previous Terrapin Module. The output layer is then connected either to a subsequent Terrapin Module for additional processing or to the final output layer for conversion to metasurface commands. Each CNN includes a 1D convolution, a batch normalization, and a rectified linear unit activation function. The 2nd level CNNs have kernel lengths of 5, 17, and 33 to increase the receptive field by 125 kHz, 500 kHz, and 1 MHz, respectively. A 1D max pooling layer with pool size of 4 is included to provide a pooling window of 125 kHz as well. The quantity $N_{xx}$ indicates a tunable parameter for the number of convolutional filters at each branch and stage, acting as a dense or fully connected layer for the global correlations. The convolutions with unit length kernels serve to buffer and condition the inputs to each stage, and the single layer 1st branch maintains the receptive field sizes from previous modules. The outputs of each branch are concatenated together to form the module output, preserving the receptive field sizes for subsequent layers. }
\end{figure}

From previous work with the cavity, we found that the mean mode spacing is $\sim$125 kHz and we demonstrated the ability to generate strong nulls over a 500 kHz bandwidth \cite{frazierWavefrontShaping2020}. This suggests we should use a pooling window of 125 kHz and allow the receptive field to increase by 125 kHz and 500 kHz at each stage, or layer in the deep learning network. After experimenting, we found that adding a 5th stage which increased the receptive field by 1 MHz helped to further improve performance. We refer to the final version of our module as a "Terrapin Module", a block diagram of which is shown in \ref{fig:modified_inception_module}.

\section{Offline Training Results for 3 x 3 Binning}
\label{app:3x3}
\begin{figure*}
\includegraphics[width = 17.2cm]{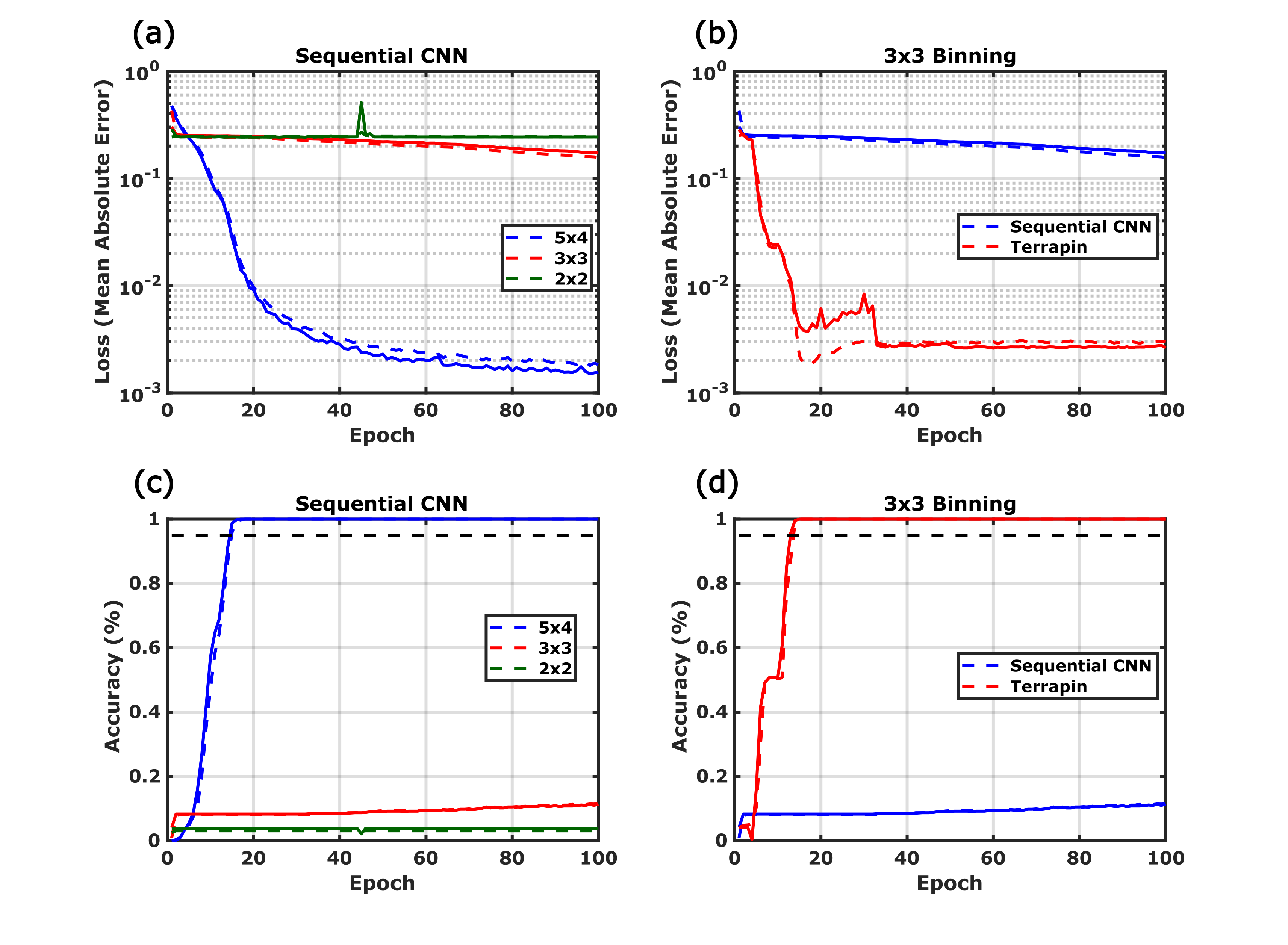}
\caption{\label{fig:cnn_vs_terrapin} \bf Sequential neural network performance with complex scattering systems. \normalfont The solid lines indicate results for the validation data while the dashed lines indicate results for the training data. \bf a) \normalfont and \bf c) \normalfont Evolution of the loss function and accuracy for the $5\times4$, $3\times3$, and $2\times2$ binning cases using the sequential CNN model. Only the $5\times4$ binning case is able to significantly reduce the loss function and provide reasonable accuracy. There is no separation between the validation results and the training results, indicating that there is not an issue with too little data. \bf b) \normalfont and \bf d) \normalfont Evolution of the loss function and accuracy for the $3\times3$ binning case using the sequential CNN and Terrapin Modules. The Terrapin Module provides similar loss and accuracy to the $5\times4$ binning case with the sequential CNN.}
\end{figure*}

For the $3\times3$ binning configuration, the purely sequential network did not perform very well and was unable to learn the relationships for either the training or validation sets. This inspired the modified inception module that we defined as the Terrapin Module in the main paper. The performance difference between the sequential CNN model and the Terrapin Module is shown in Fig. \ref{fig:cnn_vs_terrapin}, which presents training results for the $5\times4$, $3\times3$, and $2\times2$ binning cases. The sequential CNN is not able to train very well for the more complicated systems ($3\times$ and $2\times2$ binning cases), while the Terrapin Module is able to exploit the more complicated relationships and provide similar performance to the sequential CNN model on the $5\times4$ binning case. 

\begin{figure*}
\includegraphics[width = 17.2cm]{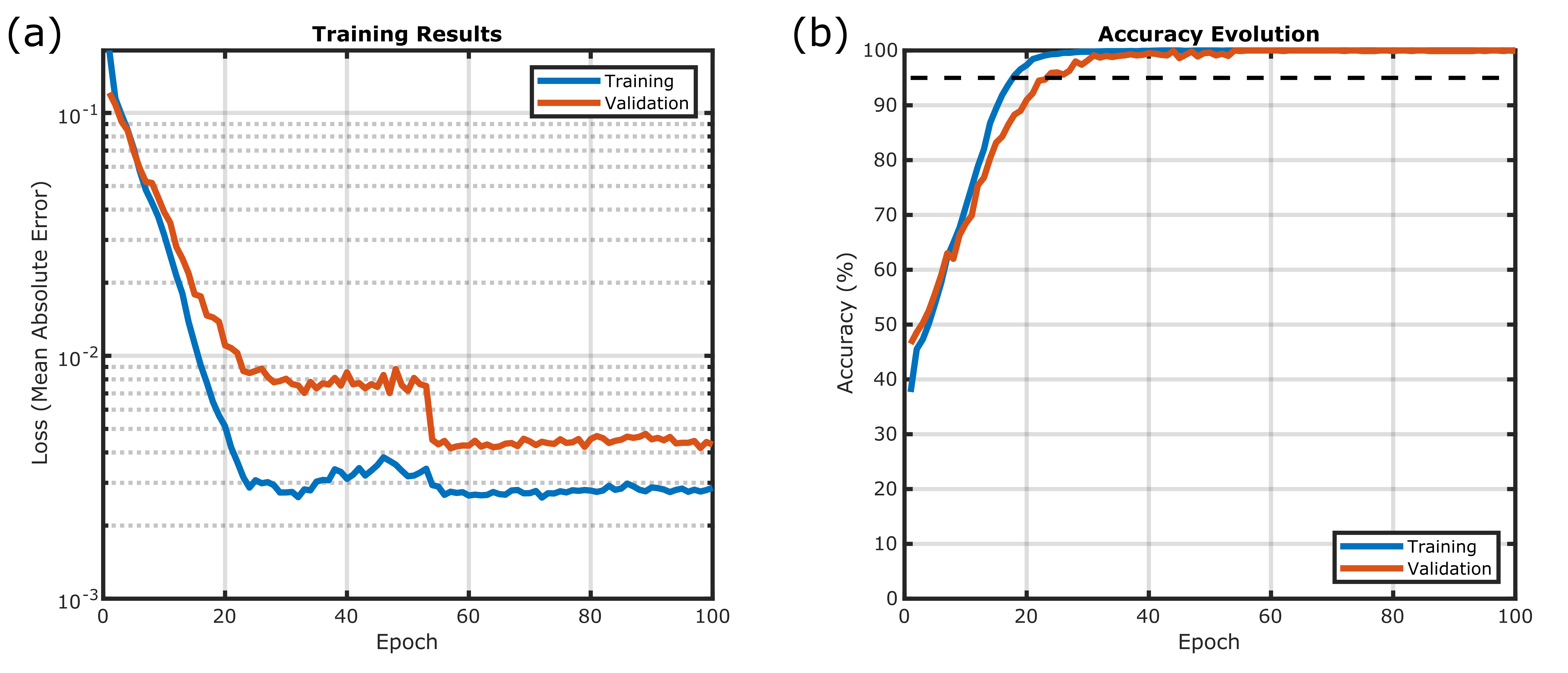}
\caption{\label{fig:performance_3x3_binning} \bf Deep learning performance with complex-valued layers for 3x3 Binning. \normalfont \bf (a) \normalfont Evolution of the loss function for the training and validation sets over 100 epochs. The loss function hits a plateau at approximately Epoch 33 but shows an additional drop at Epoch 54 when the learning rate is reduced. \bf (b) \normalfont Evolution of the accuracy for the training and validation sets over 100 epochs. Accuracy provides the relative number of sets of commands that were predicted without error, and shows that perfect prediction was achieved on both the training and validation sets in less than 100 epochs. The loss function continues decreasing after the accuracy saturates at 100\% because it is continuous and evaluated on the floating point predicted values and the decrease indicates the network is still learning and improving its estimate.}
\end{figure*}

The results for the $3\times 3$ binning case with complex-valued layers are shown in Fig. \ref{fig:performance_3x3_binning}. The impact of reducing the learning rate on a plateau can be seen at Epoch 54, where a drop in the learning rate by a factor of 10 induces a drop in the loss function of approximately a factor of 2.

\section{Offline Training Results for 2 x 2 Binning}
\label{app:2x2}
For the $2\times2$ configuration with 4000 sets of data, we were able to achieve $>$98\% accuracy on the training set, but were limited to $\sim$50\% accuracy on the validation set. The discrepancy between training and validation results is a hallmark of overtraining. In this particular case, the validation results were improving but stalled as the training results approached 100\% accuracy. The error landscape became extremely small with a negligible gradient, so there was no direction to take and continue learning. The network therefore learned specific features of the training set rather than general features of the full range of possible responses. This suggests the overtraining is due to having a limited amount of data (only 4000 sets). We captured a larger amount of data (10,000 sets) and were able to achieve $>$95\% accuracy on both the training and validation sets. Perfect accuracy for the validation set may be possible with the collection of an even larger amount of data.

\section{Scattering Fidelity Loss}
\label{app:scattering_fidelity}
\begin{figure*}
\includegraphics[width = 17.2cm]{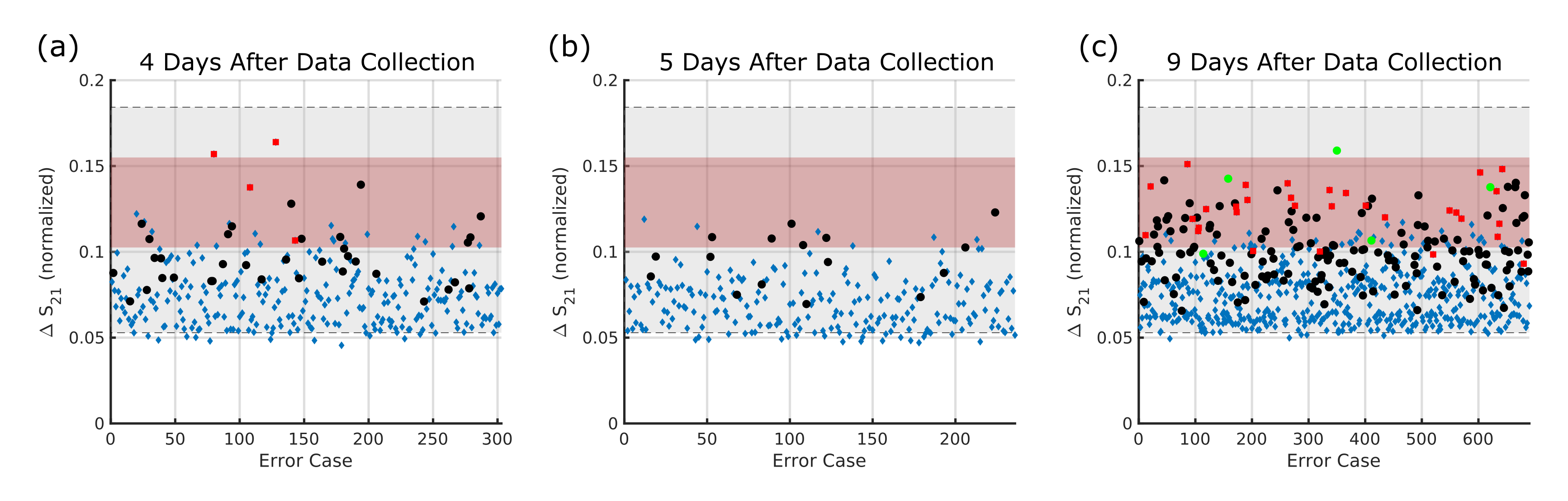}
\caption{\label{fig:scattering_fidelity} \bf Scattering fidelity loss over time.  \normalfont $\Delta S_{21}$ for online validation sets taken a specified time after the training data was collected. The shaded regions show the extent of the single element Hamming distance results from the training data. The grey region shows the full range from maximum to minimum, and the red region shows the 25$^{\text{th}}$ and 75$^{\text{th}}$ percentiles. The blue diamonds indicate cases with a single prediction error, the black circles indicate cases with 2 prediction errors, the red squares indicate cases with 3 prediction errors, and the green circles indicate cases with 4 prediction errors. These panels show that the $\Delta S_{21}$ for prediction errors is very small, and in the lower region of the statistics covered by observed cases with single element Hamming distances. \bf (a) \normalfont Validation 4 days after collecting training data, 2000 sets of commands were tested with 303 mispredictions for an accuracy of 84.9\%.  \bf (b) \normalfont Validation 5 days after collecting training data, 2000 sets of commands were tested with 236 mispredictions for an accuracy of 88.2\%. \bf (c) \normalfont Validation 9 days after collecting training data, 2000 sets of commands were tested with 690 mispredictions for an accuracy of 65.5\%. }
\end{figure*}

Figure \ref{fig:scattering_fidelity} shows the decay in scattering fidelity for online validation at the 4, 5, and 9 day marks. The accuracy is still $>$85\% after 5 days, but the number of sets with more than 1 prediction error has increased. After 9 days, the accuracy drops to 65.5\% and many cases with 2, 3, and even 4 prediction errors are found. 

\section{Performance vs. Reverberation Time}
\label{app:reverberation_time}
\begin{figure*}[h!]
\includegraphics[width=17.2cm]{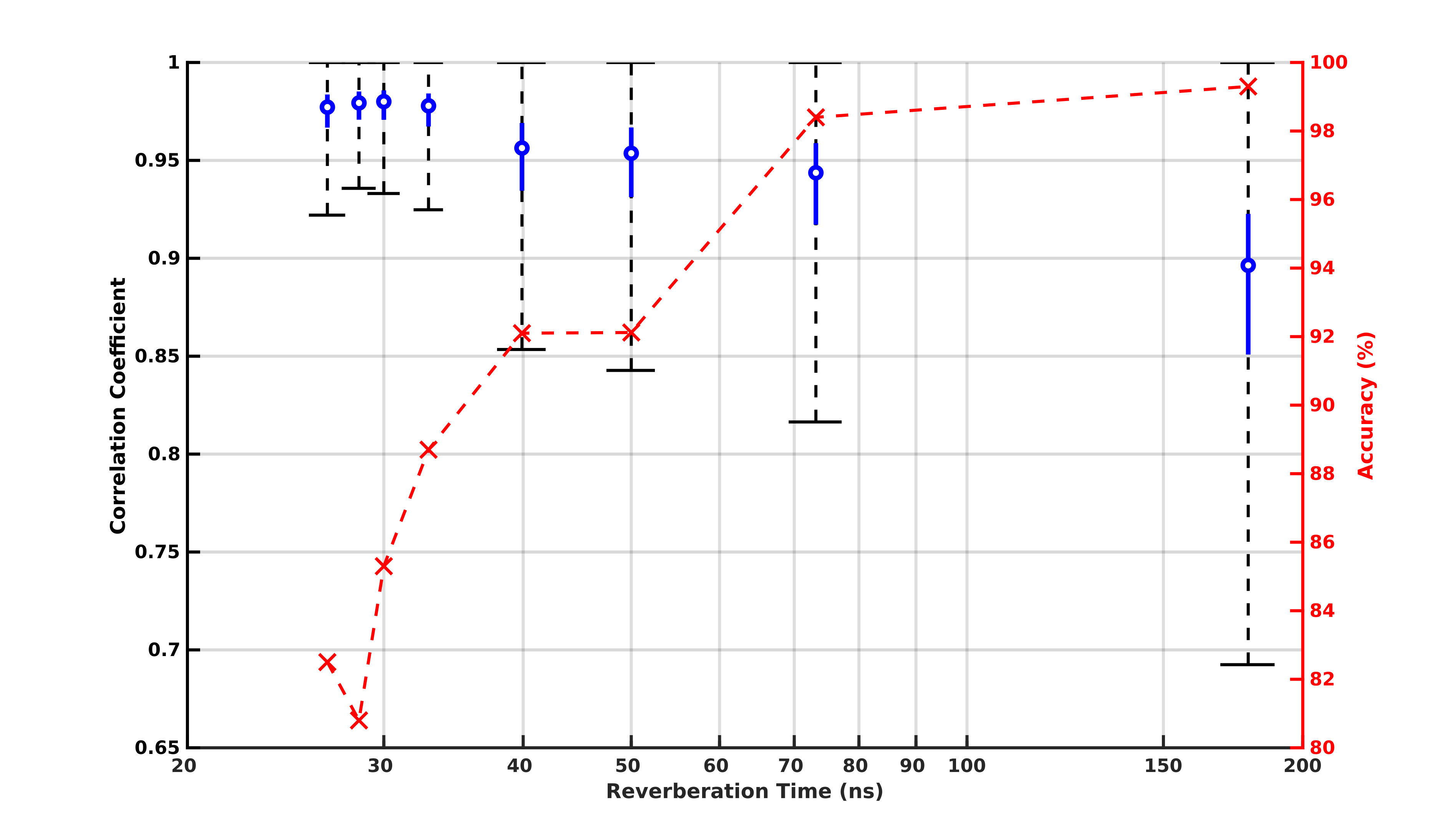}
\caption{
{\bf Deep learning performance vs. cavity reverberation time.} 
Performance of the deep learning network for different cavity loss configurations as specified by the cavity reverberation time ($x$-axis). The reverberation time is shown on a log scale to highlight the behavior for high loss configurations (short reverberation times). The statistics of the correlation coefficient over the $\sim$50 million combinations of measurement sets are shown relative to the left-hand $y$-axis. The blue line shows the extent between the quartiles, the blue circle indicates the median value, and the dashed black line shows the full extent. The achieved accuracy of the deep learning network is shown as the dashed red line relative to the right-hand $y$-axis, with individual points represented by a cross. The trend shows an inverse relationship between the correlation coefficient and the achieved accuracy of the deep learning network, indicating that the deep learning network struggles to identify features in the data when it is highly correlated.}
\label{fig:reverb}
\end{figure*}

An additional set of experiments was performed to determine the impact of cavity reverberation time on the performance of the deep learning network. To increase the losses in the cavity (and decrease the reverberation time), RF absorbent materials were placed inside the cavity. For each loss configuration, an ensemble of measurements was collected using the mechanical mode stirrer and the reverberation time was estimated from the power delay profile (PDP) \cite{hollowayEarlyTimeBehavior2012}. The mode stirrer was then set to a fixed position and another ensemble was collected for training data with 10,000 random metasurface configurations (following a biased random coin toss approach). The correlation coefficient was computed over all possible measurement pairs, for $~5\times10^7$ combinations, to assess how highly correlated the training sets were. The deep learning network was then trained using the same network and parameters as previously discussed and the results are shown in Fig.~\ref{fig:reverb}. The cavity reverberation time ranged from 23 ns to 179 ns. The statistics of the correlation coefficients are shown relative to the left-hand axis, and show the median value, quartiles, and full extent. The achieved accuracy of the deep learning network on the training set is shown as the dashed red line relative to the right-hand axis and indicates that the accuracy and correlation coefficients are inversely related. The deep learning network is capable of operating in extremely complicated scattering environments, but the performance degrades as the cavity losses increase. This is because ray trajectories do not persist as long for high loss systems; the number of bounces for a given trajectory is reduced, which means there are fewer rays intercepted by the metasurface.

\section{Future Directions}
\label{app:future_directions}
Future directions will refine our technique to intentionally scramble (or unscramble) waves propagating through a complex scattering environment. Three aspirational goals include: 1) tuning the scattering responses through a controller that optimizes the system for a given application at arbitrary frequencies and bandwidths. Specific metrics include minimizing transmitted power for coldspot generation, minimizing scattering matrix eigenvalue magnitudes for coherent perfect absorption, or minimizing the bit-error rate for wireless communication; 2) introducing feedback from the environment to dynamically update the controller and react to changing environmental conditions; and 3) realizing a fully autonomous systems that enables persistent and robust smart radio environments that do not require human intervention.

For on-the-fly learning and adaptation to changing environmental conditions, we propose the future use of reinforcement learning \cite{watkinschristopherjohncornishhellabyLearningDelayedRewards1989,bertsekasReinforcementLearningOptimal2019}, which is at the intersection of artificial intelligence and optimal control. Reinforcement learning uses an agent that interacts with an environment to learn about it and then manipulate that environment in order to maximize (minimize) a reward (cost function), leading to the development of optimal control policies. In particular, the subset of reinforcement learning known as deep or double deep ``Q'' learning is gaining traction as a method for controlling quantum states \cite{bukovReinforcementLearningDifferent2018,mackeprangReinforcementLearningApproach2020,wautersReinforcementlearningassistedQuantumOptimization2020}. Deep ``Q'' learning uses a deep learning network to estimate a quality matrix that scores the result of taking a particular action, while double deep ``Q'' learning uses two estimates to limit the implementation of poor control policies from overestimation \cite{vanhasseltDeepReinforcementLearning2015}. The deep learning network architecture developed in this paper is well suited for estimation of this quality matrix. 

Specification of an arbitrary scattering condition in the current implementation is cumbersome, as the complete $S_{21}$ response over the full 3-4 GHz measurement window must be defined. For practical engineering applications, we prefer a simpler method of defining a desired wave scattering condition. Deep reinforcement learning also helps in this case, as it scores the performance of an agent through a scalar, positive, and real-valued metric. The agent uses the deep learning network to learn the relationship between metasurface commands and $S_{21}$ responses, but the complicated details are hidden from the user. There are therefore 2 learning components to deep reinforcement learning: an inner deep learning network that learns how to map $S_{21}$ responses onto metasurface commands, and an outer agent based loop that learns how to use the inner deep learning network to optimize the desired metric. This metric can be the total power in a specified bandwidth for cold spot generation, the magnitude of the eigenvalues of the full $S$-matrix at a given frequency for coherent perfect absorption, or the bit error rate for communications systems.

Learning from scratch can be slow and may not be fast enough to adapt to changing environmental conditions. In this case, transfer learning, or using information about a similar problem to accelerate training for another one, can be incorporated into the reinforcement learning strategy \cite{taylorTransferLearningReinforcement}.


\end{document}